\documentclass[namedreferences]{solarphysics}
\usepackage[hyperref,optionalrh,showbiblabels]{spr-sola-addons} 
\usepackage{graphicx}                    
\usepackage{courier}                    
\usepackage{amssymb}                    
\usepackage{color}                       
\usepackage{url}                         

\usepackage[hyperref,optionalrh,showbiblabels]{spr-sola-addons} 
\usepackage{enumerate}
\usepackage{multirow}
\usepackage{tabularx}
\usepackage{upgreek}

\newcommand{\adv}{    {\it Adv. Space Res.}} 
 
\newcommand{\aap}{    {\it Astron. Astrophys.}}

\newcommand{\aj}{     {\it Astron. J.}} 
\newcommand{\apj}{    {\it Astrophys. J.}}
\newcommand{\apjl}{   {\it Astrophys. J. Lett.}}
\newcommand{\apss}{   {\it Astrophys. Space Sci.}}

\newcommand{\mnras}{  {\it Mon. Not. Roy. Astron. Soc.}}

\newcommand{\solphys}{{\it Solar Phys.}}
\newcommand{\sovast}{ {\it Sov. Astron.}} 
\newcommand{\ssr}{    {\it Space Sci. Rev.}}
\newcommand{\gp}{    {\it GeoPhysics}}
\newcommand{\arfm}{    {\it Ann. Rev. Flu. Mech.}}
\newcommand{\npg}{{\it Nonlinear Proc. Geophy.}}
\newcommand{\na}{{\it New Astron.}}
\newcommand{\NI}{{\it Neuro. Image}}
\newcommand{\prl}{{\it Phys. Rev. Lett.}}
\newcommand{\planss}{{\it Planet. Space Sci.}}
\newcommand{\apjs}{{\it Astrophys. J. Supp.}}
\newcommand{\zap}{{\it Zeits. Astrophys.}}
\newcommand{\jrasc}{{\it J. Roy. Astron. Soc. Can.}}
\chardef\us=`\_

\begin{document}

\begin{article}

\begin{opening}

\title{Solar-Cycle Characteristics in Kodaikanal Sunspot Area: North--South Asymmetry, Phase Distribution and Gnevyshev Gap}


\author[corref,email={ravindra@iiap.res.in}]{\inits{B}\fnm{B.}~\lnm{Ravindra}\orcid{https://orcid.org/0000-0003-2165-3388}}
\author[email={parthares@gmail.com,partha240@yahoo.co.in}]{\inits{Partha}\fnm{Partha}~\lnm{Chowdhury}\orcid{}}
\author[email={jj@iiap.res.in,jdotjavaraiah@gmail.com}]{\inits{J}\fnm{J.}~\lnm{Javaraiah}\orcid{}}
\address{Indian Insititute of Astrophysics, Koramangala, Bangalore 560034, INDIA}
\address{University College of Science and Technology, Department of Chemical Technology, University of Calcutta, 92, A.P.C. Road, Kolkata, 700009, West Bengal, India}
\address{No. 58, 5th Cross, Bikasipura, Bengaluru-560 078. Formerly with Indian Institute of Astrophysics, Koramangala, Bengaluru - 560034, India}


%

%
\runningauthor{B. Ravindra et~al.}
\runningtitle{Solar--Cycle Characteristics in Kodaikanal Sunspot area data}


\begin{abstract}
The solar activity is asymmetric in both hemispheres in almost all cycles. This asymmetry is observed both in cycle amplitude and period. We have used about 90~years of sunspot-area data from the Kodaikanal Solar Observatory to study the North--South asymmetry in sunspot activity. The monthly mean sunspot-area showed the northern hemisphere dominated in Solar Cycles 16, 19, and 20, and the southern hemisphere dominated in Cycles 18, 22, and 23. The 13-months smoothed data indicated that in Cycle 17 and 21, the northern and southern hemisphere showed equal amplitude. Cumulative sunspot area showed that the northern hemisphere dominated in Cycles 18, 19, 20, and 21, with a large difference between the two hemispheres in Cycles 19 and 20. The northern hemisphere activity led by 12, 15, and 2 months in Cycles 20, 21, and 22, respectively. No significant phase difference is found between the two hemispheres in Cycles 16, 17, 18, 19, and 23. The wavelet technique is used to find Rieger-type periodicities in the sunspot cycles. The cross-wavelet analysis of these data sets showed several statistically significant common periodicities like the Rieger-type periodicities and Quasi-biennial oscillations. The Gnevyshev gap was found in both the hemispheric data in Cycles 16, 18, 21, 22, and 23. These results are consistent with the earlier reported characteristics of 
North--South asymmetry in sunspot-area data. These results suggest that the Kodaikanal Observatory data complement the existing sunspot data from other observatories to study solar activity over long and short periods.     
\end{abstract}

%

\end{opening}

%
\section{Introduction}\label{sec:intro}
Sunspots are dark and are most easily visible features on the Sun, even with a moderate-sized telescope. Most solar-activity phenomena are connected to sunspots.
The sunspot record goes back to the time of Galileo \citep{2007AdSpR..40..929V}. However, the daily record of the sunspot observations started in 1874. These observations started in several observatories around the world at different times \citep{2013A&A...550A..19R,2020Ap&SS.365...14R}, and until now, the sunspot-area and number have the longest record compared to any other data available in history \citep{2014SSRv..186..105E,2017yCat..36010109U}. This longest record of data is useful to understand solar irradiance, space climate, the Earth's atmospheric chemistry, and rain pattern \citep{2007LRSP....4....2H,2010RvGeo..48.4001G,2017JGRA..122.3888Y}.

An eleven-year periodicity is observed in the sunspot activity. Solar dynamo believed to be operating at the base of the convection zone is assumed to be responsible for the observed solar activity on the Sun \citep{2017shin.confE.171C}. There are several dynamo models proposed starting from the flux-transport dynamo model of Babcock--Leighton \citep{1991ApJ...375..761W,1995A&A...303L..29C,1999ApJ...518..508D} to 3D MHD dynamo models \citep{2017ApJ...847...69K}. All are able to describe the 11-year periodicity with uncertainity in their cycle amplitudes \citep{2013SSRv..176..289J, 2014A&A...563A..18P}.

Solar cycles show the dominance of one of the hemispheres compared to the other \citep{2014SSRv..186..251N}. The north--south asymmetry in the sunspot-area is reported as small during the maximum period compared to the minimum period \citep{1977SoPh...52...53R}. It is reported that sometimes during the minimum period of the cycle, large sunspot groups appear in one of the hemispheres, close to the Equator, that could cause the imbalance in the N--S asymmetry in the area and magnetic flux. It is also found that the northern and southern hemisphere peak at a different time of the activity cycle. This phase difference between the two hemispheres is part of asymmetry in their activity \citep{1957ZA.....43..149W, 1971SoPh...20..332W}. According to dynamo theory, the stochastic fluctuations in the dynamo mechanism can lead to the asymmetry in the two hemispheres \citep{1992A&A...253..277C, 2007PhRvL..98m1103C}.

It is now well accepted that the N--S asymmetry exists in a variety of solar activity indices, but most or all of them are related to the sunspot activity. The asymmetry in the activity indices is reported in sunspot-area data having more than 145 years \citep{1990A&A...229..540V, 1993A&A...274..497C, 2013ApJ...768..188C, 2015AJ....150...74Z, 2017A&A...603A.109B}. It is also observed in sunspot-number data as well as in Joy's law for a few solar cycles \citep{2006A&A...447..735T, 2011NewA...16..357B, 2013SoPh..287..215M, 2017A&A...599A.131L, 2019SoPh..294..142C}. Apart from these, the asymmetry is reported in plage index \citep{2007ASPC..368..527D, 2019AdSpR..63.3738S}, magnetic flux \citep{2005A&A...438.1067K, 2007SoPh..246..445M, 2013ApJ...763...23S, 2004SoPh..221..151V, 2018SoPh..293..158V}, filaments or prominences observed \citep{2001SoPh..199..211D, 2010NewA...15..346L, 2015ApJS..221...33H}, differential rotation \citep{1997SoPh..170..389J, 2005SoPh..227...27G, 2018ApJ...855...84X}, etc. It is also reported in energetic events such as X-ray and H$_{\alpha}$ flare index  \citep{1996SoPh..166..201A, 2013Ap&SS.343...27D, 2013AN....334..217D, 2017JSWSC...7A..34D, 2019MNRAS.488..111D}, coronal mass ejections, radio bursts \citep{1987SoPh..114..185V, 2009MNRAS.400.1383G, 2014SoPh..289.2283G}, energetic-proton events \citep{2016SoPh..291.2117R}, coronal-hole area \citep{2015A&A...583A.127N} solar wind \citep{2015JGRA..120.3283T,2020ApJ...894...13Z}, coronal green-line index \citep{2010SoPh..261..321S}, etc. Some of the properties of N--S asymmetry can be used to understand the observed periodicities in cycles, and North--South asymmetric properties of low-latitude sunspot area can be used to predict the amplitude of the next cycle \citep{2007MNRAS.377L..34J}.

The sunspot-area data are available from the Royal Greenwich Observatory (RGO: \opencite{2015LRSP...12....4H}), the longest data set. Most of the N--S asymmetry related studies were carried out using this data set. The next available data set is from the Debrecen Observatory \citep{2016SoPh..291.3081B}, which is about 62~years long. Studies on N -- S asymmetry have also been  done using this data set \citep{2019SoPh..294...64J}. Apart from these, such studies are done using the Mt. Wilson white-light data set \citep{1984ApJ...283..373H,2010A&A...518A...7D, 2015MNRAS.447.1857B}, Michelson Doppler Imager (MDI: \opencite{1995SoPh..162..129S}) white-light image data set \citep{2009SoPh..260....5W,2015NewA...39...55R}. Another observatory that has the longest white-light data series is from the Kodaikanal Solar Observatory (KOSO: \opencite{2010ASSP...19...12H}). The data have been obtained from a single telescope since 1905, and it has collected more than 110 years of data. The data have been digitized in the past, and several studies on solar activity are made \citep{1999SoPh..188..225G, 2003SoPh..214...65S, 2010SoPh..266..247S}.

In the present study, we use the KOSO digitized data to study the N--S asymmetry in the sunspot group area for eight solar cycles. These data make up another large data set that can be used independently to study several properties of solar cycles \citep{2017A&A...601A.106M, 2020ApJ...891..151H}. In the next section, we describe the data and analysis methods adopted. In Section 3, we present the results of the study, and we end the paper with a summary and discussion. 
                      
\section{The White-light Images and Method of Analysis}\label{sec:cda}
\subsection{Kodaikanal Sunspot-Area Data}

In the present study, we have used the daily white-light sunspot observations, from 1921 to 2011. These images were recorded with the help of a 6-inch lens mounted on an equatorial-mount telescope installed at the KOSO. The images of the sunspots were recorded on lantern photographic plates starting in 1904 \citep{1993SoPh..146...27S} using the aforementioned telescope and continued until today without any change of its optical set-up since 1918 \citep{2013A&A...550A..19R, 2017A&A...601A.106M}. However, from 1975, these photographic plates were replaced with high-contrast films. These century-long sunspot data were digitized and calibrated. To extract the different features of sunspots, a semi-automated algorithm version of the STARA code \citep{2009SoPh..260....5W} was used. The area occupied by the sunspots (in the units of a millionth of a hemisphere: $\mu$Hem) and the information corresponding to their latitude and longitude measurements was extracted. This has been done for each observation starting from 1921, and the measured area was corrected for the projection effect (e.g. \opencite{2013A&A...550A..19R, 2017A&A...601A.106M}). The sunspot-area during this time span was averaged over a month and we obtained the time series of the monthly averaged sunspot-area in both the hemispheres separately. The detailed explanation of the images, whole calibration process, including data reduction and digitization of this time series was described by \cite{2013A&A...550A..19R}. The Kodaikanal data archive of digitized white-light images is hosted at kso.iiap.res.in/new/white$\_$light. 


\begin{figure}[!h]
\begin{center}
\includegraphics[width=0.45\textwidth]{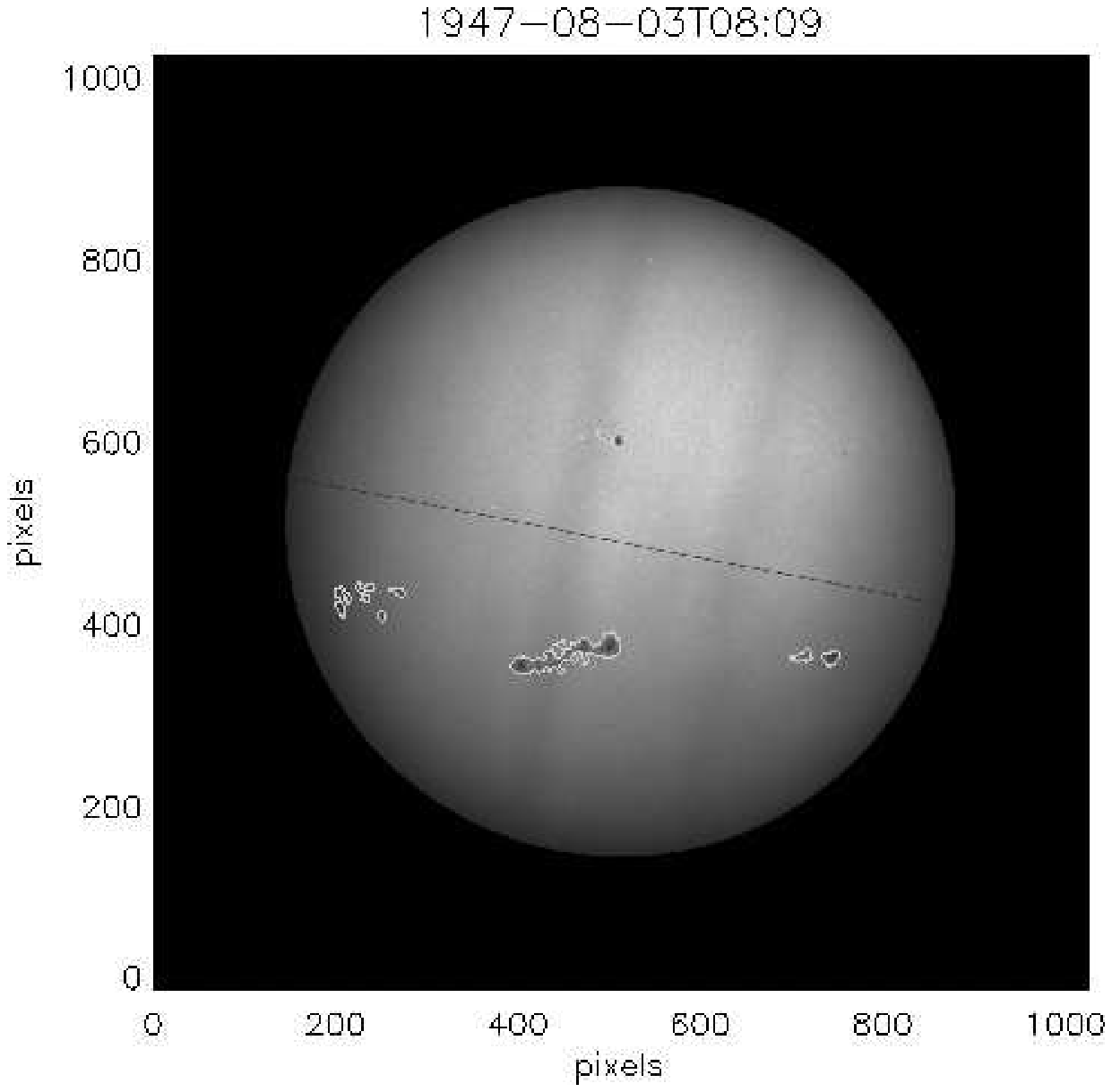} 
\includegraphics[width=0.45\textwidth]{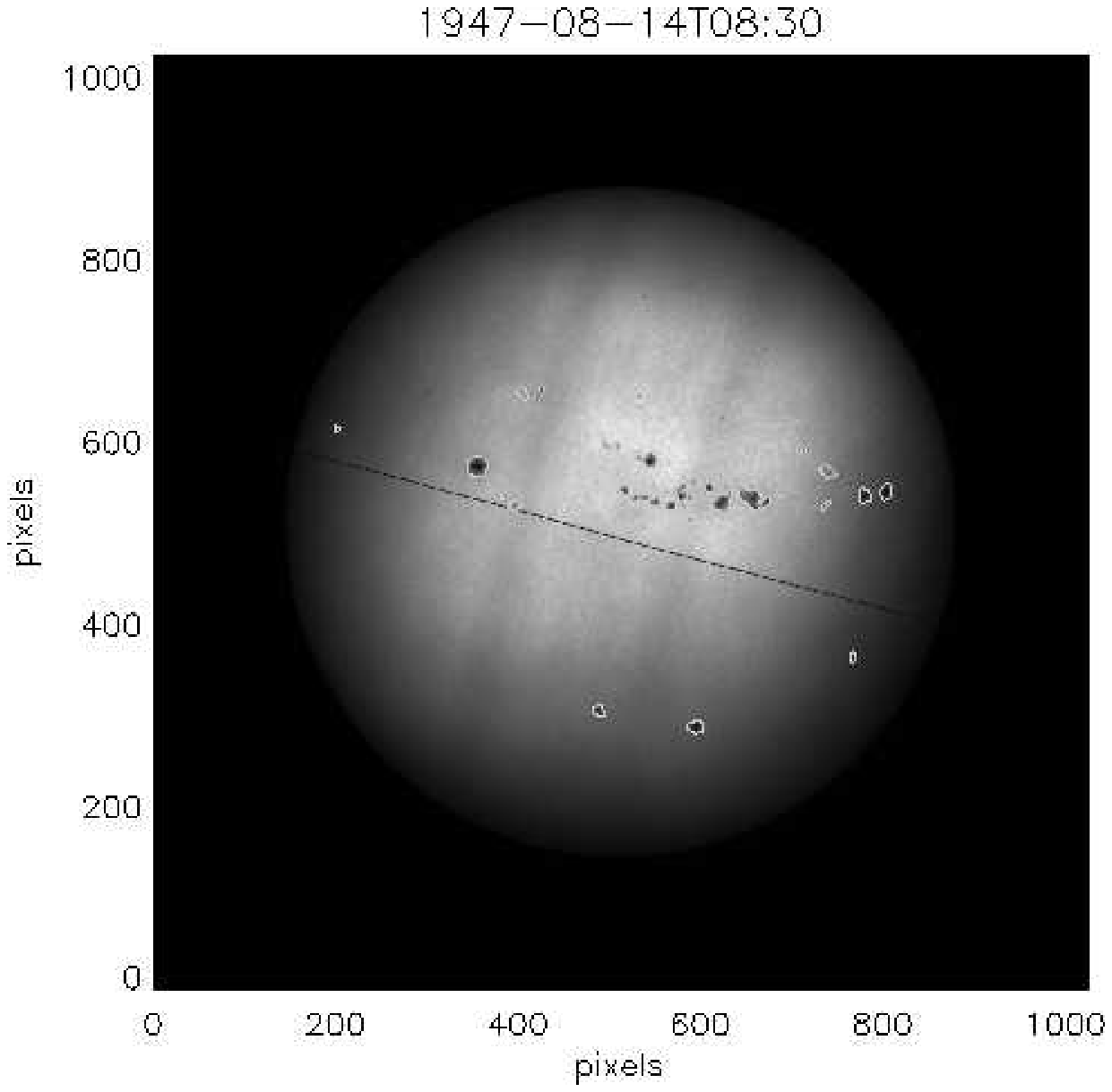} \\
\caption{The Sun's image taken on 03 August (left) and 14 (right) 1947 at the Kodaikanal Observatory. The detected sunspots in the white-light digitized images are shown with contours.}
\end{center}
\label{fig:1}
\end{figure}

After digitization and calibration following \cite{2013A&A...550A..19R}, the sunspots are detected. Figure~1 depicts the two solar images taken on 03 and 14 August 1947, at the Kodaikanal Solar Observatory with the detected sunspots marked with contours on white-light images. 

\subsection{Methods for Kodaikanal Sunspot Area Analysis}

\begin{figure}[!h]
\begin{center}
\includegraphics[width=0.45\textwidth]{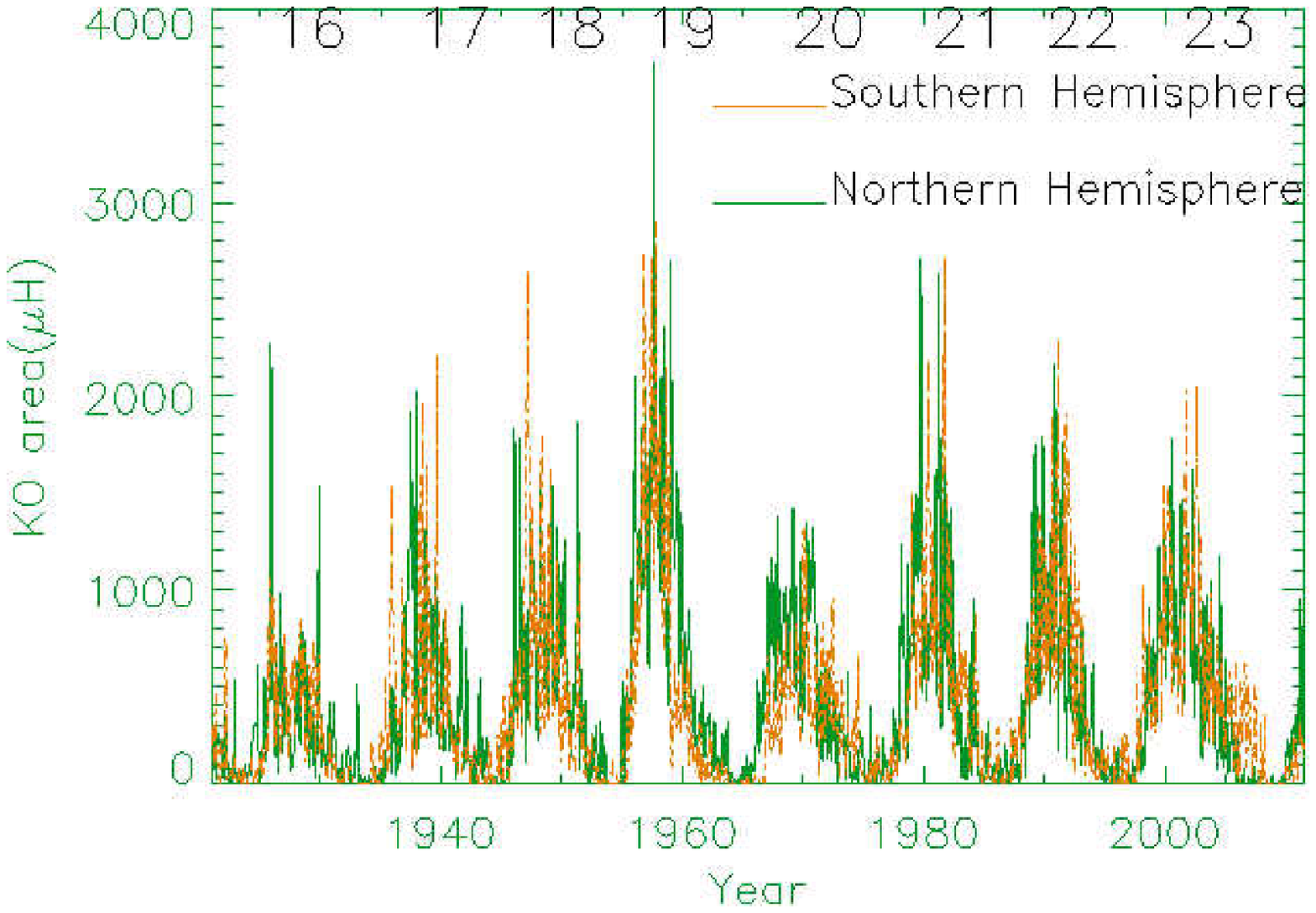} 
\includegraphics[width=0.45\textwidth]{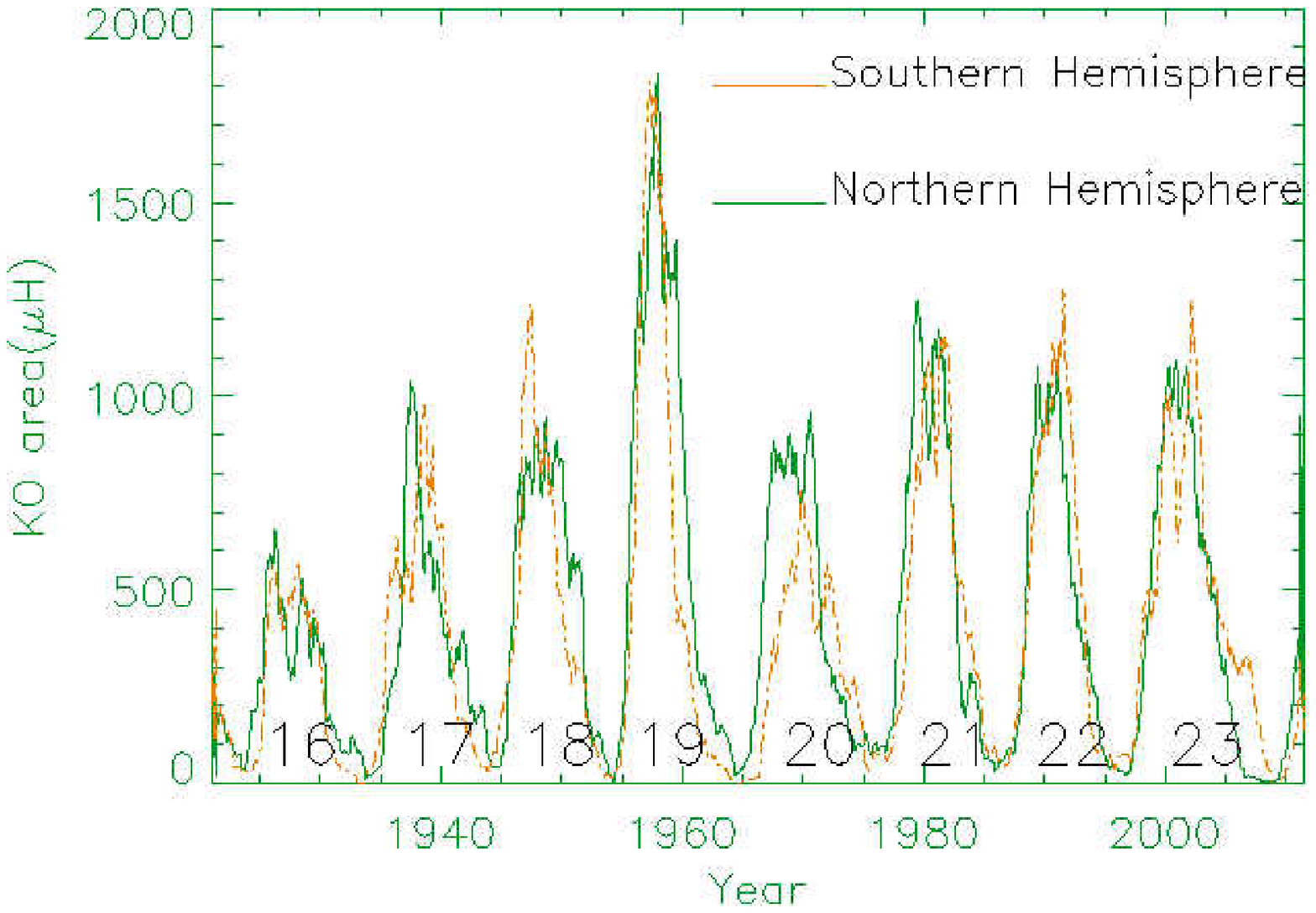} \\
\caption{Left: The monthly mean KOSO sunspot-area of the opposite hemispheres is plotted from 1921 to 2011. Right: The same time series with 13-month smoothed data.} 
\end{center}
\label{fig:2}
\end{figure}

Figure~2(a) shows the monthly averaged sunspot-area of both the hemispheres. The 13-month smoothed sunspot-area plotted against time is shown in Figure~2(b). Both figures indicate a regular cyclic pattern of about 11 years, and the figure also exhibits the characteristic property of each solar cycle, that is different from others. 
We have quantified the sunspot-area asymmetry between the opposite hemispheres by using the definition of the normalized north--south asymmetry index 
[$A_\mathrm{a}$ = ( $A_\mathrm{n}$ -- $A_\mathrm{s}$ ) / ( $A_\mathrm{n}$ + $A_\mathrm{s}$ )],	
where  $A_{\mathrm{n}}$ and $A_{\mathrm{s}}$ represent the monthly mean northern and southern hemispheric sunspot area respectively. We have made a plot of the temporal evolution of $A_{\mathrm{a}}$ data set for each of the Solar Cycles 16 -- 23 separately. Then, by looking at the trend, we have plotted the oblique straight line. This is to examine the change in the trend of asymmetry in the course of any solar cycle under study.

Solar activities are found to have a phase shift or time lag/lead between the northern and southern hemispheres in a sunspot cycle \citep{2009SoPh..255..169L, 2012MNRAS.419.3624M, 2019ApJ...875...90L}. To investigate the time lead or lag between the opposite hemispheres, we have calculated the cross-correlation (CC) between the 13~months smoothed monthly mean 
sunspot-area data of northern and southern hemispheres. This is done for each of Cycles 16 -- 23.

We have also determined the periodic variations of the asymmetry in time series using the spectral decomposition technique of the wavelet-analysis method. However, to detect periodicities in sunspot-area asymmetry, we have used monthly values of north--south
difference of area $\Delta$=$A_{\mathrm{n}}$ -- $A_{\mathrm{s}}$. This is because 
there is some
confusion about the existence of an $\approx$11-year periodicity in the relative
asymmetry: $(A_{\mathrm{n}} - A_{\mathrm{s}})/(A_{\mathrm{n}} + A_{\mathrm{s}})$ \citep{1997SoPh..170..389J, 2005A&A...431L...5B, 2020SoPh..295....8J}. \cite{1992JRASC..86...89Y} for the first time suspected the 11-year periodicity in the relative asymmetry might be an artifact of the numerator. However, we have also applied the wavelet technique on the relative asymmetry time series to examine the similarity and differences between the two time sequences.

The wavelet technique \citep{1992ARFM...24...395F} is a useful tool for identifying oscillatory components within a given signal or time series, which converts a 1D signal into a 2D signal in the time--frequency domain and also provide information about the exact location of the detected periods. We applied a complex Morlet wavelet \citep{1982GP...47...203M} function to both absolute and relative asymmetry time series as, 
\begin{equation}
\psi_n(\eta) = \pi^{-1/4}\mathrm{e}^{\mathrm{i}\omega_{0} n}\mathrm{e}^{-\eta^2}/2
\end{equation}

Here, $\omega_{0}$ is a non-dimensional frequency and we have adopted $\omega_{0}$ = 12 for mid-term frequency (low periodic zone) range and $\omega_{0}$ = 6 for the low-frequency range (long periods) \citep{1998BAMS..79..61T, 2005A&A...438.1067K, 2019SoPh..294..142C}. The wavelet spectrum suffers from edge effects as wavelet power decreases by a factor $\mathrm{e}^{-2}$ at the edges \citep{2004NPG..11..561G}, which gives rise to a cone-of-influence (COI). This COI is plotted with a bold dashed line to represent the edge effects. The thin black contours show the periods above 95\,\% confidence level inside the COI, assuming red-noise background and detected using the recipe by \cite{2004NPG..11..561G}. ``Global Wavelet Power Spectra'' (GWPS) are computed for all of these local wavelet plots, representing wavelet power at each period scaled and averaged over time. The working principle of this method is similar to that of Fourier power spectra, and the confidence level (95\,\%) of the GWPS plots are determined using the method provided by \cite{1998BAMS..79..61T}. 


To investigate the phase shift and common quasi-periods between the monthly sunspot-area data of the northern and southern hemisphere, we have employed the cross-wavelet transform (XWT) and wavelet coherence (WTC) techniques. For two series $X_{n}$ and $Y_{n}$, the quantity $\mid$  $W^{XY}_{n}$ $\mid$ shows the amount of common power between the two time series in the XWT spectrum as a function of time and frequency \citep{2004NPG..11..561G}. The phase relation between the two time series is represented by arrows in the XWT spectrum, with the following convention: arrows pointing right are in-phase; pointing left are in anti-phase; pointing 
up are the second series leads by 90$^{\circ}$; pointing down are the first series leads by 90$^{\circ}$. A random distribution of the arrows represents phase mixing between the two time series under study. The wavelet transform coherence (WTC) between two time series $X_{n}$ and $Y_{n}$ is defined as

\begin{equation}
R^{2}_{n} = \frac{\mid S(s^{-1}W^{XY}_{n}(s))\mid^{2}}{S(s^{-1}\mid W^{X}_{n}(s)\mid^{2}) S(s^{-1}\mid W^{Y}_{n}(s)\mid^{2})}.
\end{equation}

Here, $S$ is a smoothing operator, considered as a localized correlation coefficient in 
time--frequency space, and its value lies between 0 and 1. WTC measures the cross-correlation between two time series, even though the common power is low in the XWT spectrum \citep{2004NPG..11..505M, 2004NPG..11..561G, 2010NI..50..98C}. To draw both the XWT and WTC plots using the Morlet wavelet technique under the red-noise approximation, we have utilized the MATLAB code developed by \cite{2004NPG..11..561G}, which is available at  grinsted.github.io/wavelet-coherence/. Finally, we have investigated the presence of the Gnevyshev Gap (GG: \opencite{1963SvA.....7..311G,1967SoPh....1..107G}), in both N and S hemispheres separately using monthly averaged sunspot-area time series. We used a methodology similar to the one described by \cite{2010SoPh..261..193N} to identify the GG in each solar cycle.

\section{Results}
\subsection{N--S Asymmetry and its Temporal Variations}

Figure~2(a) shows the temporal variation of the monthly averaged sunspot area of both the N and S hemispheres for the time span of 1921 to 2011, covering complete Solar Cycles 16 -- 23 and a small part of the rising branch of Cycle 24. \cite{2017A&A...601A.106M} mentioned that in the KOSO area data the amplitude of Cycle 23 is large in KOSO data compared to the RGO data. Similarly, the Cycle 18 amplitude is smaller than the RGO measured amplitude. In order to balance this, we identified the dates where the data is not present in the KOSO data set in Cycles 18 and 23. If the gap between the consecutive observing days is more than two days then we filled those data points from the RGO data. The resulting plot is shown in Figure~2(a). This plot shows that the nature and amplitude of each cycle are different from others. It is clearly seen that sunspot-area data sets used in this study show about 11-year solar-activity cycle with some differences. Similar results are found in the 13-month smoothed 
sunspot-area plot (Figure~2(b)). It is observed that in the course of a sunspot cycle, the peak in the sunspot-area occurs in different time intervals with different amplitudes in the opposite hemispheres. Figure~2(b) shows that the sunspot-area possesses maximum and minimum amplitude during Cycles 19 and 16, respectively, and also shows a double peak during the cycle maximum. The southern hemisphere is dominant over the North in Cycles 18, 22, and 23
and this difference is marginal for Cycles 16 and 19. From Figure~2(a), we find that the odd-numbered cycles like 17, 19, and 21 have higher amplitudes compared to the preceding even-numbered cycles like 16, 18, and 20. This is in line with the odd--even cycle rule or Gnevyshev--Ohl rule \citep{1968P&SS...16.1311G}. 

\begin{figure}[!h]
\begin{center}
\includegraphics[width=0.7\textwidth]{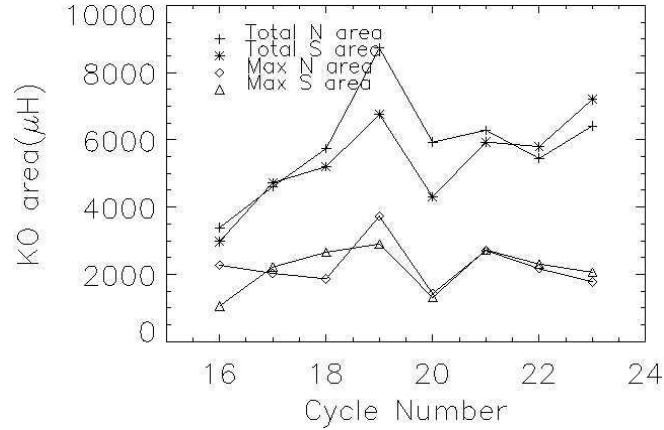} \\
\caption{The total area occupied by the sunspots in the northern (plus ($+$)) and southern (star ($\ast$)) hemisphere over each cycle from 16\,--\,23.  The value of the total sunspot-area is divided by 10 to show the other curves in the same plot. The diamond ($\diamond$) and  triangle($\triangle$) symbols represent the maximum area occupied by the monthly averaged sunspot-area in each cycle in both the northern and southern hemispheres respectively.} 
\end{center}
\label{fig:3}
\end{figure}
 
Figure~3 shows the maximum amplitude of sunspot-area in both hemispheres during the maximum of the eight solar cycles under study. From the plot, we found that the sunspot areas in both the northern and southern hemispheres achieved maximum and minimum values during Cycles 19 and 20, respectively. Solar Cycle 16 also showed the weakest value. Total sunspot-area in the southern hemisphere showed a significant rise from Cycle 20 to 21 and maintained nearly the same trend in Cycle 22 with a jump in Cycle 23, again. On the other hand, total sunspot-area in the northern hemisphere exhibited a steady rise from a Cycle 20 to 21 with a drop of its value in Cycle 22. During Cycles 17, 20, 21, 22, and 23, maximum sunspot-area occupied by the opposite hemispheres was nearly the same.     

\begin{figure}[!h]
\begin{center}
\includegraphics[width=0.45\textwidth]{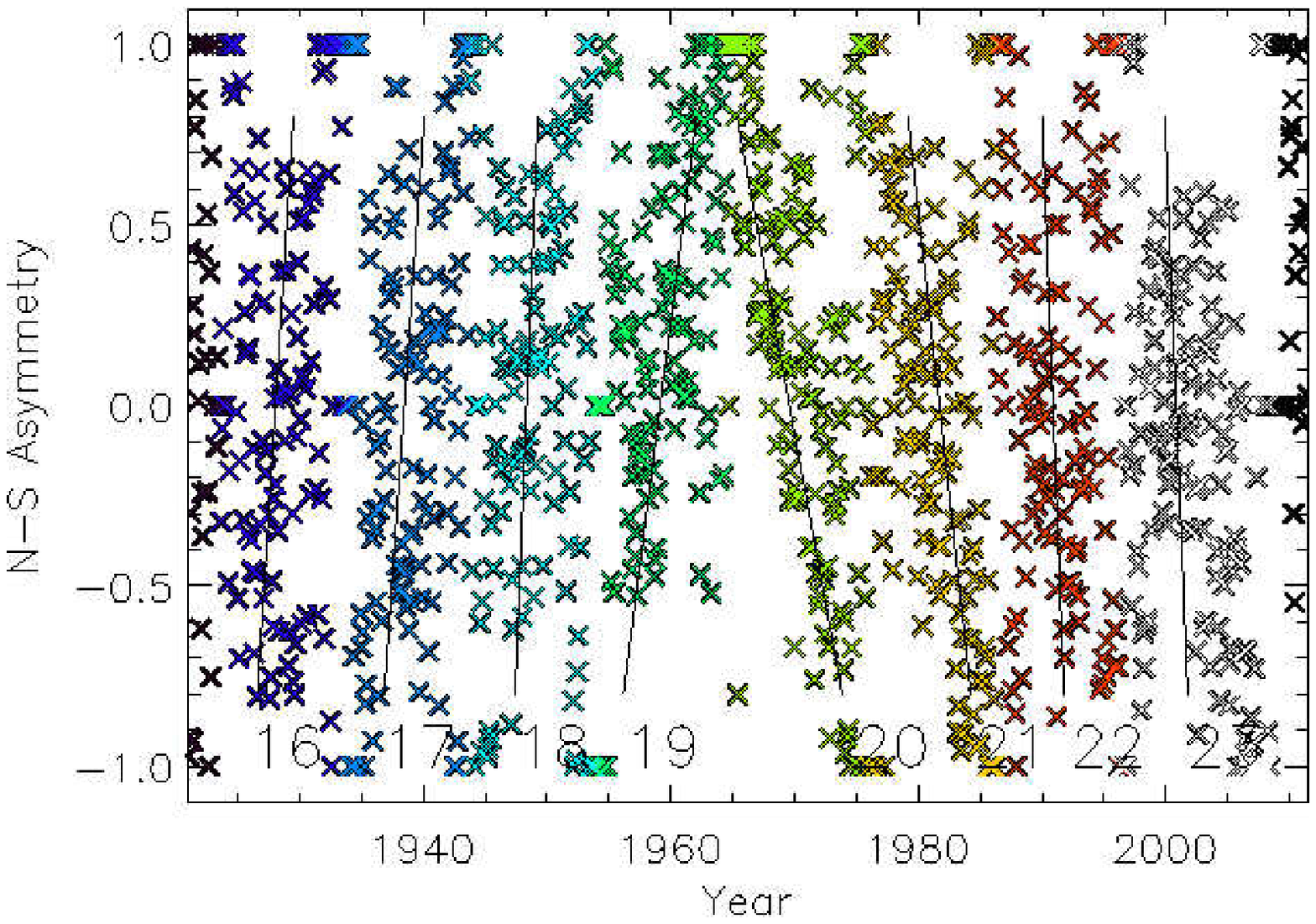} 
\includegraphics[width=0.45\textwidth]{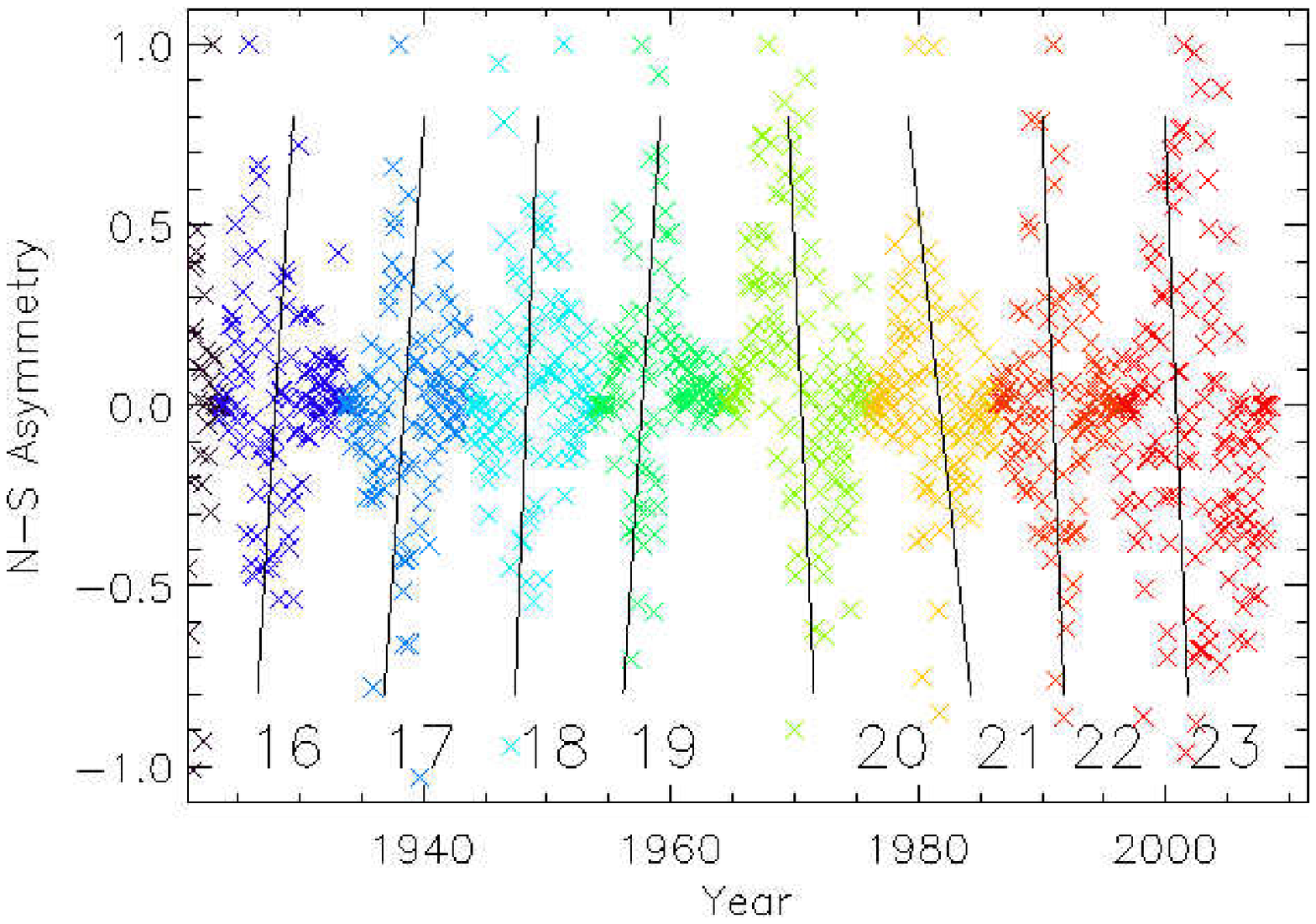} \\
\caption{Left: The relative north--south asymmetry in monthly mean sunspot areas is plotted against time. The solid lines indicate the trends during each cycle. Right: Same as the left side but for the absolute north--south asymmetry.}
\end{center}
\label{fig:4}
\end{figure}

Figure~4(left) shows the temporal evolution of the relative N--S asymmetry index $A_{\mathrm{a}}$ in the monthly averaged sunspot area for each Solar Cycle, 16 to 23. It has been found that unlike during the minimum period of the cycles, the area asymmetry values lie between --0.5 and  $+$0.5. However, around the minimum and the onset period of the solar cycles, the sunspot-area asymmetry index has somewhat large values, close to $+$1 or --1 depending on which hemisphere dominates over the other. Further, this plot shows that the slope of the line changes in each cycle, indicating some long-term (about eight cycle period) systematic change in dominance of the sunspot-area activity from one hemisphere to another.
The right hand side plot in Figure~4(right) shows the temporal evolution of the absolute asymmetry. The data points are normalized by the maximum value in the cycle series. As for the 
relative-asymmetry, also found the eight-cycle period. We have only eight and a half cycles of data, and the pattern of the inclination of the lines indicates that this could be an eight-cycle period.

\subsection{Dominant Hemispheres}

Following \cite{2007A&A...476..951C}, we applied Student's t-test to determine the dominant hemisphere in each solar cycle under study and the results are shown in Table~\ref{tab:1}.  We consider preselected error probability in the t-distribution as $<$~0.025, i.e. the difference between northern and southern hemispheric sunspot-area is statistically significant at a 95\,\% level. Otherwise, the difference in sunspot areas in two hemispheres for any cycle should be regarded as marginal. Such types of solar cycles are marked with ``--'' in the table.

\begin{table}[!h]
\caption{Cumulative sum of the monthly averaged northern and southern hemisphere sunspot-area data, probability of finding the asymmetry by chance, asymmetry index, and dominant hemispheres in sunspot-area for Cycles 16 to 23.}

\label{tab:1}
\begin{tabular}{llllll}     
	
	\hline
	Cycle No. & SSA N-hemi & SSA S-hemi & Probability & Asymmetry & Dominant hemisphere \\
	\hline
16 & 33920 & 29829 & 1.16 $\times$10$^{-1}$ & 0.064  & -- \\
17 & 46061 & 47234 & 4.15 $\times$10$^{-1}$ & -- 0.013 & -- \\
18 & 57417 & 52043 & 1.65 $\times$10$^{-1}$ & 0.049 & -- \\
19 & 87263 & 67575 & 2.01 $\times$10$^{-3}$ & 0.127 & N \\
20 & 59256 & 42950 & 2.04 $\times$10$^{-4}$ & 0.159 & N \\
21 & 62863 & 59076 & 2.61 $\times$10$^{-1}$ & 0.031 & -- \\
22 & 54349 & 57989 & 2.55 $\times$10$^{-1}$ & -- 0.032 & -- \\
23 & 64156 & 72056 & 5.17 $\times$10$^{-2}$ & -- 0.058 & -- \\

\hline
\end{tabular}
\end{table}

Table~\ref{tab:1} shows that the asymmetric nature is statistically significant for the northern hemisphere in Cycles 19 and 20. The southern hemisphere is preferred in Solar Cycles 17, 22, and 23, but the difference is not statistically significant. We also noticed that the normalized asymmetry is more pronounced at the beginning of a new cycle when the number of spots on the solar disk is small. 

From \cite{2017A&A...601A.106M}, it is clear that there is a difference in the cycle amplitude of the Kodaikanal data compared to the RGO data. This could be because of missing data, the contrast of the photographic plates, semi-automated detection of sunspots, and the software's ability to detect smaller sunspots in images. The RGO data have been cataloged using sunspot observations from different observatories around the world (Cape of Good Hope, Kodaikanal, and Mauritius) to supplement missing days or bad data. Thus, the RGO data are affected by different scaling to incorporate data points from various observatories. This may also lead to differences in values of KOSO data in comparison to the RGO data. 

Many of the earlier statistical studies \citep{2002A&A...383..648L} used the yearly averaged data to study the dominant hemisphere.  We have used the monthly averaged data, and the large difference in the amplitudes of the North--South shows as significant asymmetry. Some studies used the probability distribution function \citep{2002A&A...383..648L}, and some others have used the overlap of the North--South time series to identify the dominant hemisphere in the cycle \citep{2014SSRv..186..251N}. It appears that the hemispheric amplitude of the KOSO data in some cycles is different than other observatory data and hence the difference in the results of the dominant hemisphere.

\begin{figure}[!h]
\begin{center}
\includegraphics[width=0.9\textwidth]{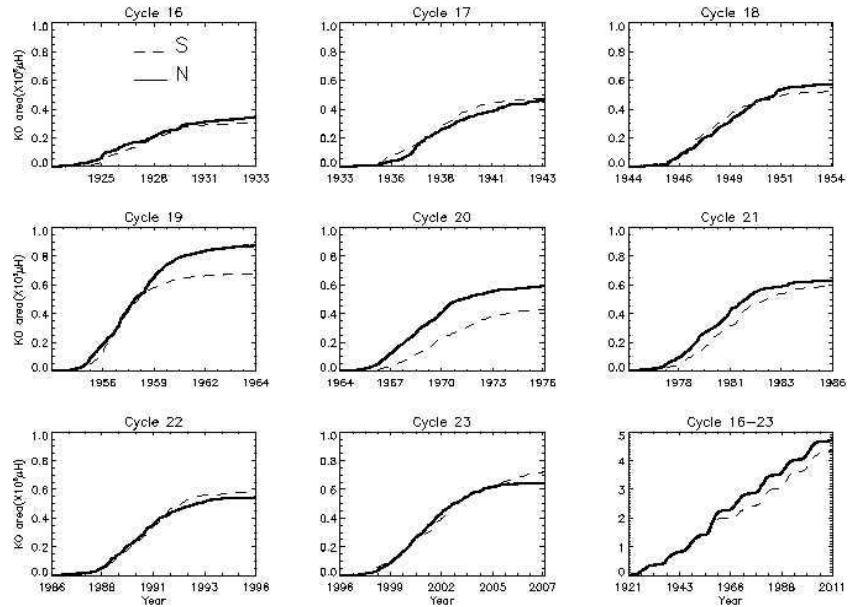} \\
\caption{Cumulative monthly counts of KOSO sunspot-areas in the northern and southern hemispheres for Cycles 16\,--\,23.}
\end{center}
\label{fig:5}
\end{figure}

Figure~5 shows the cumulative sum of the sunspot-area against time in the northern (solid line) and southern (dashed line) hemispheres for each cycle 16 -- 23 separately by utilizing the monthly averaged sunspot-area values. This indicates that Cycle 16 shows approximately equal activity levels in both the hemispheres. However, in Cycles 18, 19, 20, and 21 the northern hemisphere dominated in activity over the southern hemisphere. On the other hand, in Cycles 17, 22, and 23, the southern hemisphere is more active than the northern hemisphere. This plot also shows the time period when the dominance of activity occurs during the progress of any cycle.  It has been found that during Cycle 20 the northern hemisphere's activity is always highly dominant, which is a unique property of this cycle. During Cycles 22 and 23, although the activity of the northern hemisphere was higher than the southern one in the ascending phase, the situation was shifted to the southern one around the maximum epoch. This property of the Kodaikanal sunspot-area data set is consistent with the results previously obtained by \cite{2015AJ....150...74Z} for ESAI sunspot-area database and \cite{2019SoPh..294..142C} for sunspot-numbers. The bottom-right plot in Figure~5 shows the cumulative sum of the sunspot-area for the northern and southern hemispheres for the Cycles 16 -- 23. This plot shows that a small or vanishing asymmetry before Cycle 19, and after that, there is a dominance of the northern hemisphere is seen. This difference could be because in Cycles 19, 20, and 21 the activity was higher in the northern hemisphere than the southern hemisphere. During these cycles, the northern hemisphere is dominant. 

\begin{figure}[!h]
\begin{center}
\includegraphics[width=0.9\textwidth]{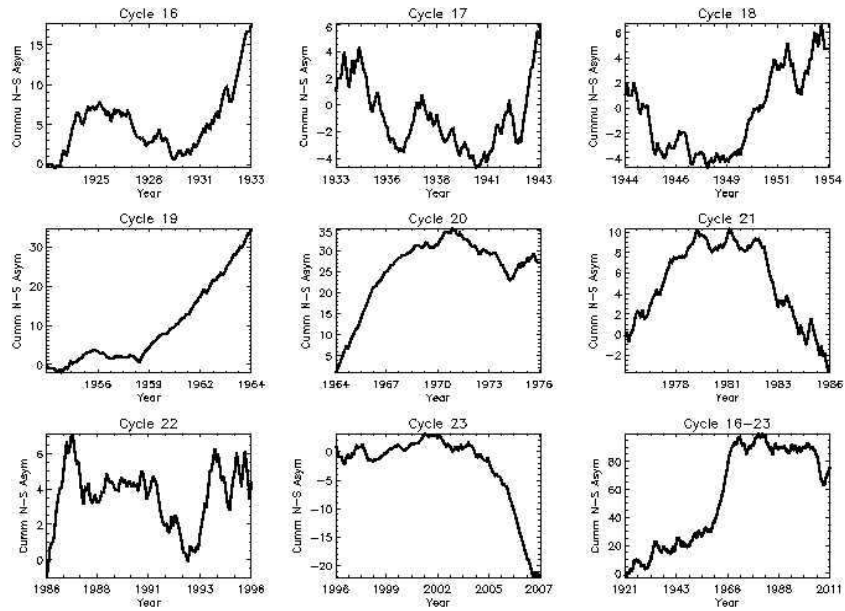} \\
\caption{The cumulative N--S asymmetry index is plotted for each Cycle, 16 -- 23. The bottom right plot is the cumulative N--S asymmetry index for all the cycles.}
\end{center}
\label{fig:6}
\end{figure}

The north--south asymmetry is small around the solar-cycle maximum phase when large numbers of spots are present in the photosphere, and it is large during the cycle minimum when a small number of spots present on the disk \citep{2009SoPh..255..169L, 2015NewA...39...55R}. Figure~6 shows the variation of cumulative asymmetry during
each cycle under investigation. As can be seen in this figure,
except for Cycle 21 all of the remaining cycles end up a with large value of the
cumulative asymmetry during the end of the cycle. The pattern of variations
of the cumulative asymmetry is considerably different from one cycle to
another. The upward and the downward trends in the values of
the cumulative asymmetry imply the dominance of the northern- and
southern-hemispheres, respectively. In some cycles (Cycles 16, 17, and 22)
there are trends of about 6\,--\,7 year periodicity. The trends in other
cycles suggest the existence of longer than a 11-year periodicity. 
The cumulative index of all the cycles (bottom right-most plot in Figure~6) shows there is steady increase in the index until the Cycle 19.  There is a sudden jump in the cumulative index during the declining phase of Cycle 19. This is consistent with the pattern seen in Figure~4, suggesting that there
could be a long periodic (about eight-cycle period) change in the N--S asymmetry.
However, we need to have more data to conclude this result firmly.   

\subsection{Periodicities in the N--S Asymmetry of Sunspot Areas}

\begin{figure}[!h]
\begin{center}
\includegraphics[width=1.0\textwidth]{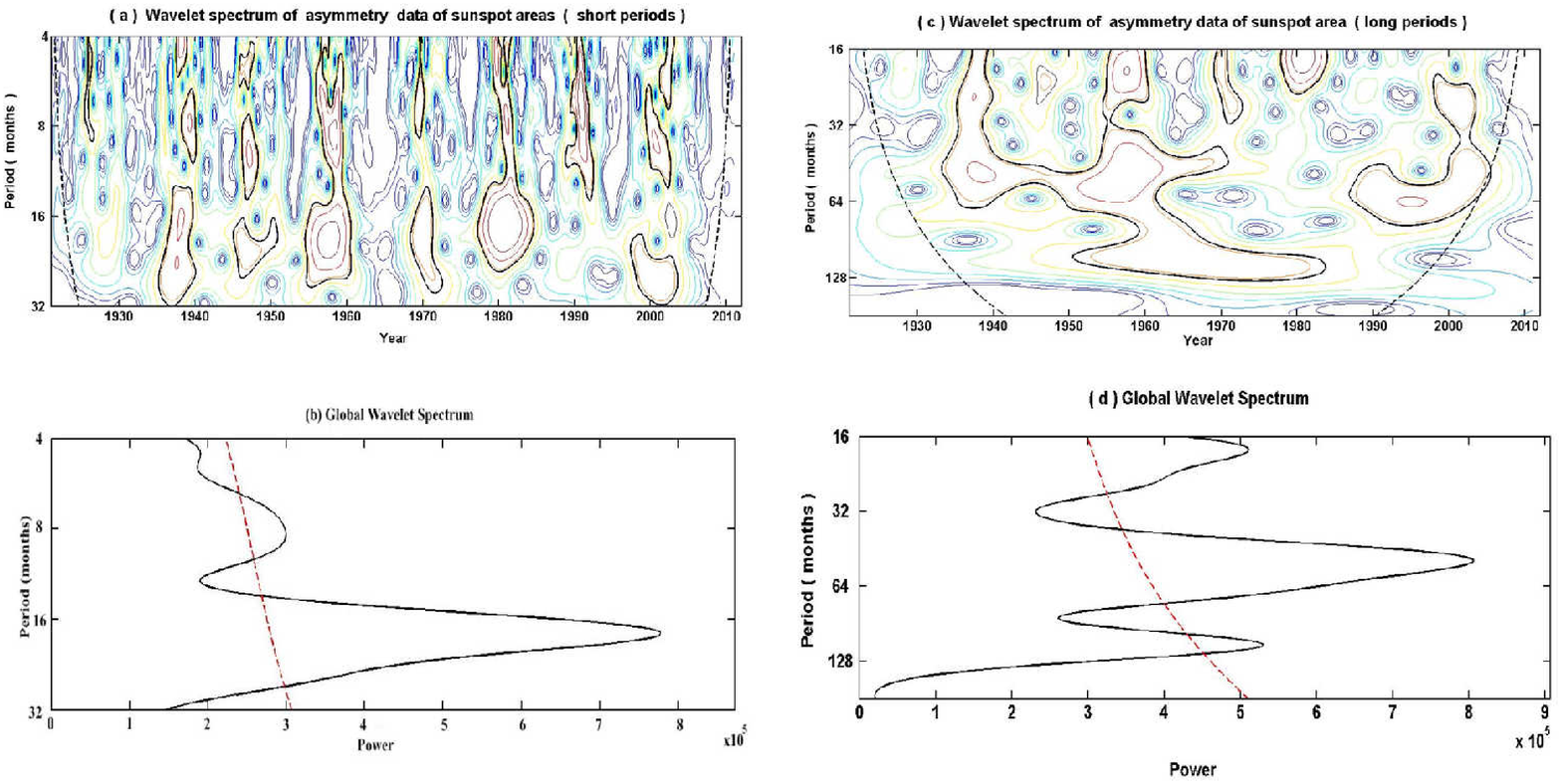} \\
\includegraphics[width=1.0\textwidth]{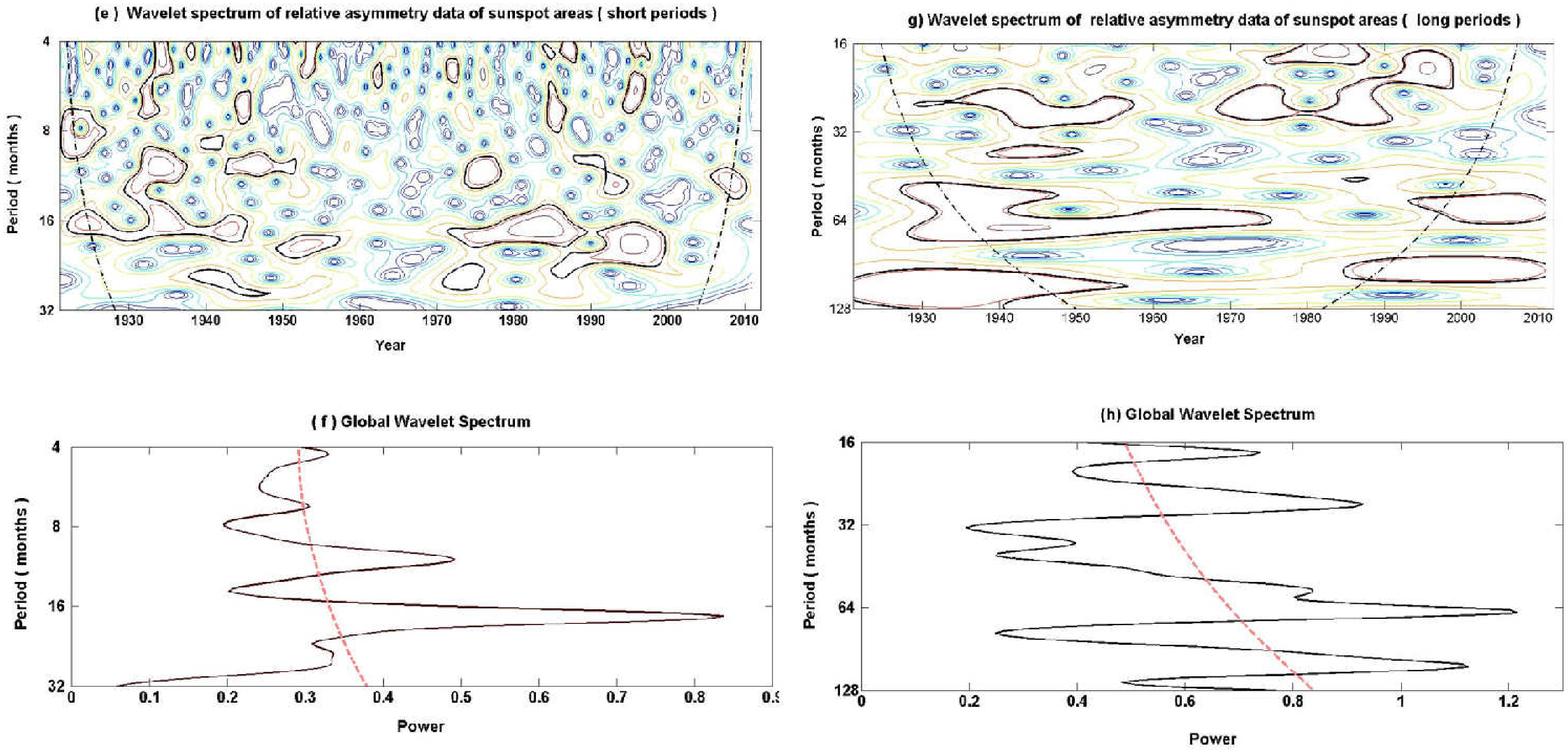} \\

\caption{Panels a and b exhibit the local and global power spectra of the monthly mean absolute-asymmetry data series from KOSO sunspot-area for the range 4\,--\,32 months. Panels c and d are similar to panels a and b for the range 16\,--\,130 months. The 95\,\% confidence level in all GWPS is indicated by red dotted lines. Panels e and f are the same as panels a and b but for the relative-asymmetry data series. Similarly, panels g and h are the same as panels c and d but for the relative-asymmetry time series.}
\end{center}
\label{fig:7}
\end{figure}

Intermediate-term quasi-periodic variations in the absolute and relative asymmetry time series of sunspot area studied using the Morlet wavelet technique and the results are displayed in Figure~7. We find several short- and mid-term quasi-periodicities during Cycles 16 -- 23 in the sunspot asymmetry data. Figure~7(a) shows that Rieger and near-Rieger type periods are present in all solar cycles including the recent Cycles 22 and 23 with a varying length between four to eight months. We have detected the presence of a quasi-annual period in Cycles 18, 19, 21, 22, and 23. Quasi-biennial oscillations (QBO) in the range of 1.3 to 2 years were present during different phases in Cycles 18, 19, 20, 21, and 23 in both of the data sets. Other long-term QBOs in the range of 2.5 to 4 years were significant approximately from 1935 to 1972, covering Cycles 17 to 20 and from 1988 to 2002, which covers Cycles 22 and 23 in absolute-asymmetry data. In relative-asymmetry data, a period in the range of 2\,--\,2.5 years was present mainly from 1935 to 1955 and from 1970 to 1990. Thus the observed Rieger-type periods and QBOs show intermittent behavior in asymmetry data. A quasi-period in the range of five years was also detected in Cycles 18 to 20 and again in Cycles 22 and 23 in 
absolute-asymmetry data. However, this period was absent in Cycle 22 for relative-asymmetry data and its presence in Cycle 23 is also weak. Along with these intermediate-term periodicities, the signatures of the 11-year period of sunspot cycle was also seen in both types of asymmetry data. However, the temporal evolution of this Schwabe-cycle period is different in each time series. The plots of GWPS (Figure~7(b), (d), (f), and (h)) of the asymmetry time series also exhibit statistically significant periodicities of Rieger type, QBOs, $\approx$~5, and $\approx$~10 years. 

\subsection{Time Delay and Phase Shift between the Hemispheres}

\begin{figure}[!h]
\begin{center}
\includegraphics[width=0.97\textwidth]{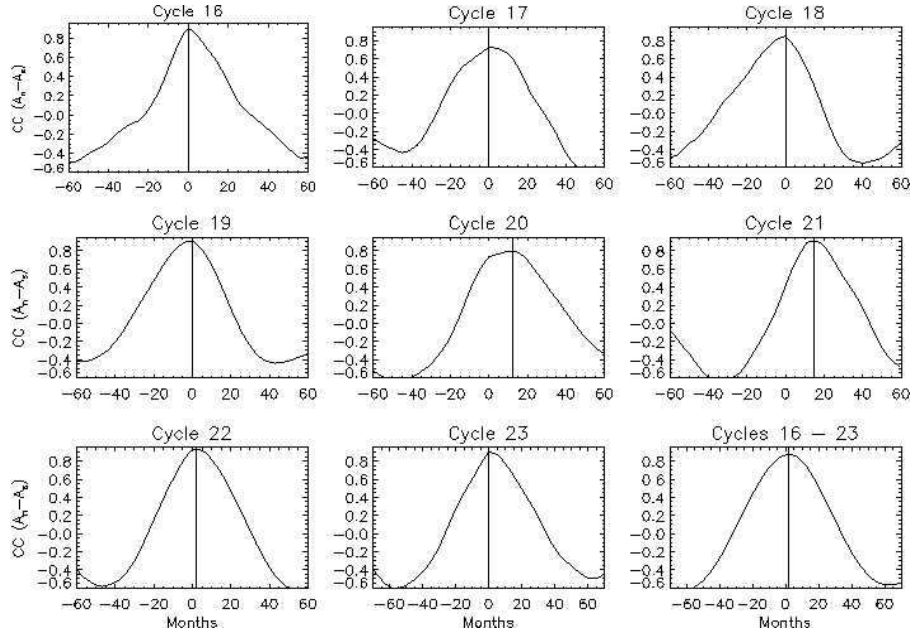} \\
\caption{The phase lag between the occurrence of peak between the northern and southern hemisphere in each cycle. The phase lag is shown in months. The solid vertical line represents the time of maximum correlation. The peaks at positive and negative values of lag indicate north and south lead respectively.}
\end{center}
\label{fig:8}
\end{figure}

We have adapted a cross-correlation technique to detect the degree of how sunspot areas in the two hemispheres are related \citep{2004ASPC..325..157H, 2016AJ....151...70D}, and the results are shown in Figure~8. These plots indicate the time lead or lag of sunspot-area time series separately for Cycles 16 to 23. The results are listed in Table~\ref{tab:2}. The bottom-right plot in the same figure shows the overall phase lead of the northern hemisphere by two months, if one considers the eight Cycles 16 -- 23. 

\begin{table}[!h]
	\caption{Time difference between the northern and southern hemisphere sunspot-area to reach their peak values and the maximum values of correlation (Max Cor) in Solar Cycles 16 -- 23. The zero means there is no time delay between the two hemispheres in reaching their maximum. The positive value of time difference indicates the lead of the northern hemispheric activity over the southern hemisphere expressed interms of months.}
	
	\label{tab:2}
	\begin{tabular}{lll}     
		
		\hline
	Solar Cycle No. &  Time lag/lead [months] & Max Cor \\
		\hline
16 & 0 & 0.90\\
17 & 0 & 0.73\\
18 & 0 & 0.85\\
19 & 0 & 0.91\\
20 & 12 & 0.80 \\
21 & 15 & 0.90\\
22 & 2 & 0.93\\
23 & 0 & 0.90\\

\hline
\end{tabular}
\end{table}

Table~\ref{tab:2} shows that in Cycle 20, 21, and 22, the northern hemispheric activity leads the southern by about 12, 15, and 2 months, respectively. However, in other cycles no delay in the activity levels is observed between the hemispheres. Compared to the earlier results \citep{2009SoPh..255..169L, 2013ApJ...765..146M} there is a difference in the phase-leading hemisphere in KOSO data. This is due to the difference in the measured area in the KOSO data compared to RGO data. Since we do not have long-term data, we can not conclude that the eight-cycle pattern exists in the hemispheric phase difference.

\begin{figure}[!h]
\begin{center}
\includegraphics[width=0.8\textwidth]{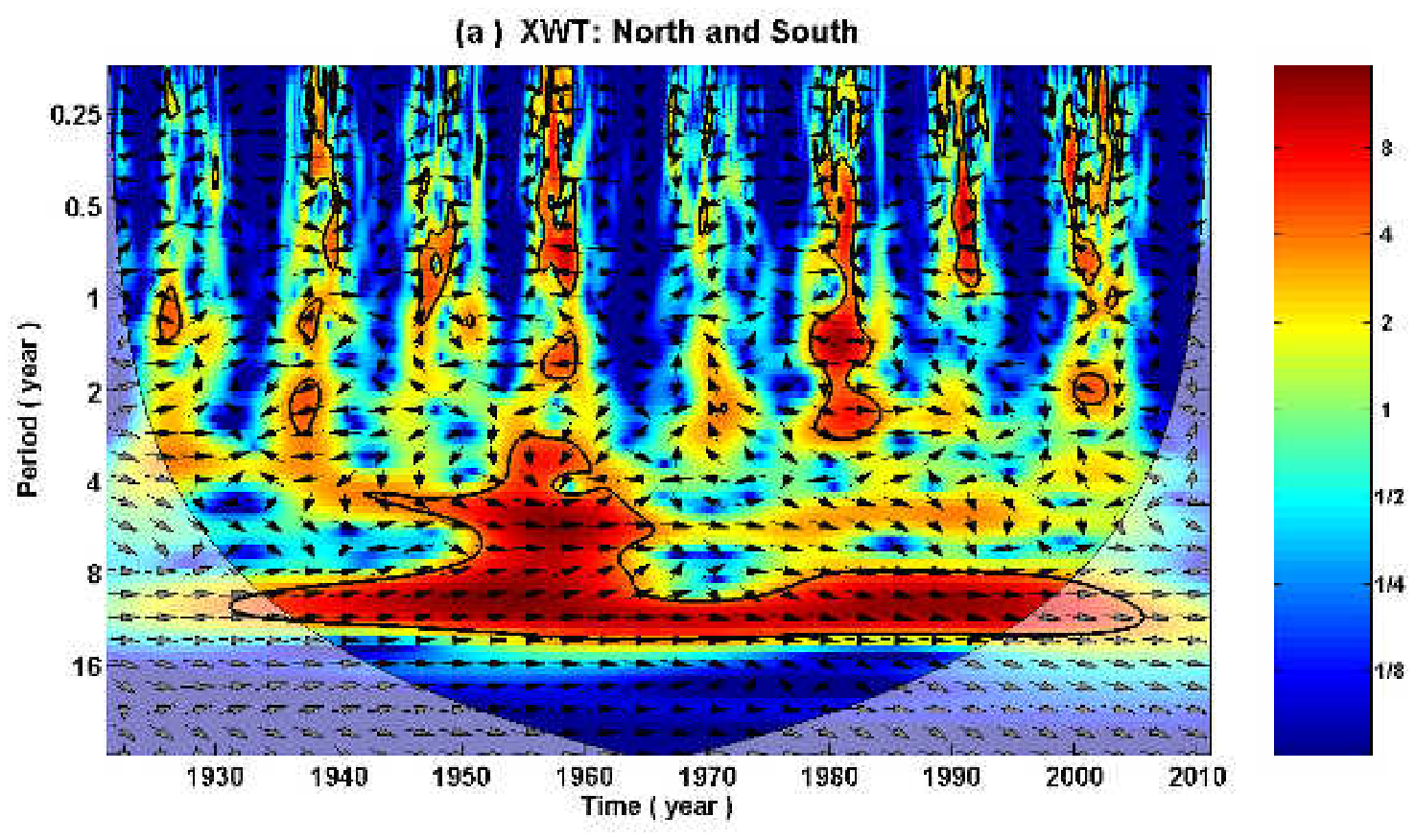}
\includegraphics[width=0.8\textwidth]{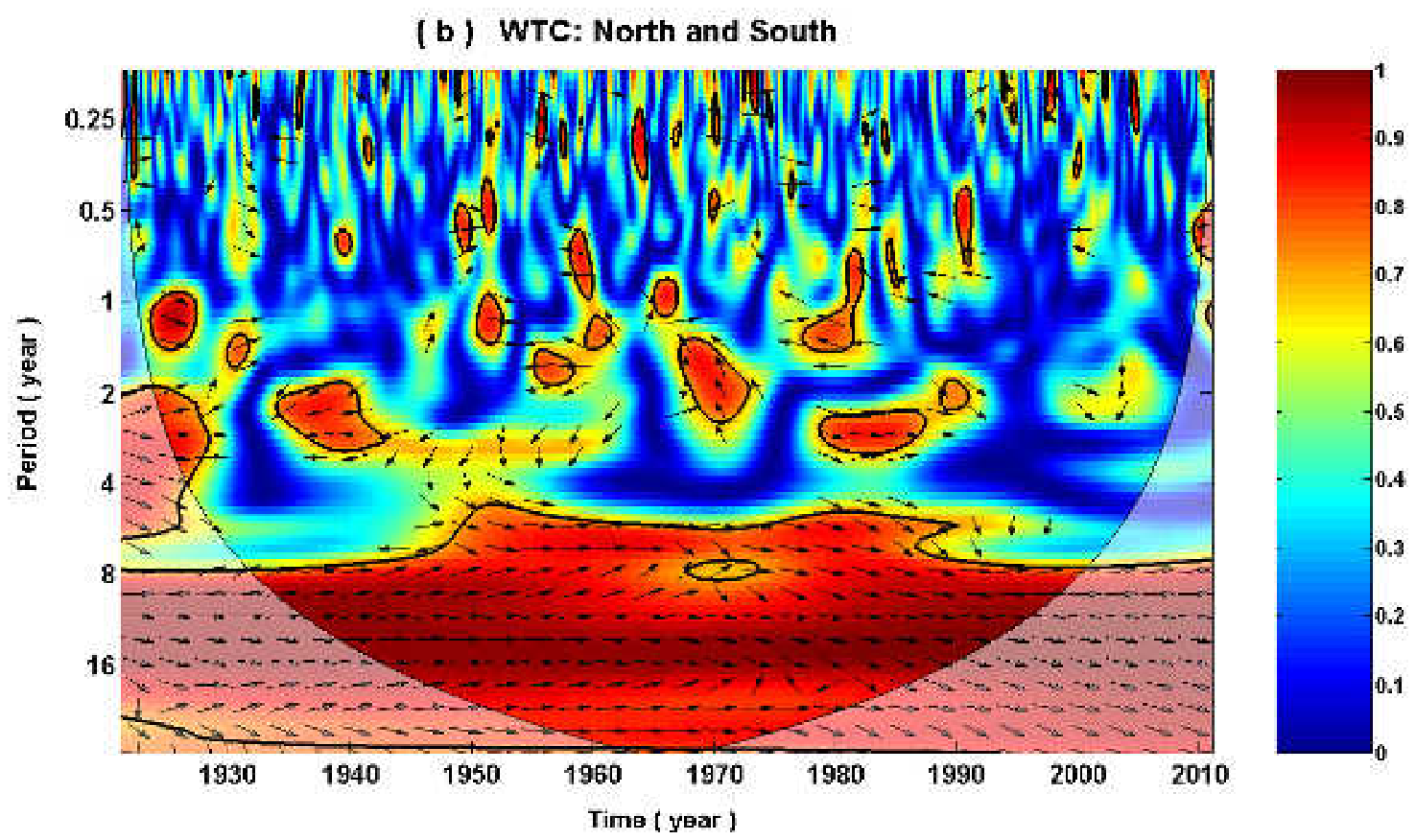}
\caption{(a) XWT and (b) WTC power spectra for the KOSO hemispheric area for Cycles 16 to 23. The solid black contours represent 95\,\% confidence level in both panels and the thin line indicates the cone of influence (COI). Phase relationship is represented by arrows.}
\end{center}
\label{fig:9}
\end{figure}

The cross-correlation method used here accounts for the total length of any solar cycle, but the phase shift may take place at any time during the cycle \citep{2010SoPh..261..193N}. Therefore, we have utilized the wavelet coherence method to realize better the relative-phase relationship and common quasi-periods of sunspot area between the opposite hemispheres \citep{2019SoPh..294..142C, 2019ApJ...874...20O, 2020SoPh..295....8J}, and the results are shown in Figure~9.  From the XWT plot, it is evident that Rieger and Rieger-type periods 
(130\,--\,190 days) are prominent and common in both hemispheres along with QBOs in the range of 1.2 to 4~years. However, the arrows are randomly distributed, indicating there exists strong phase mixing in these periods. Hence, no regular oscillatory pattern has been found in those periodic regions. The WTC spectrum reveals that this periodic zone is highly correlated.  Both of the wavelet-coherence plots indicate that within the periodic zone of 
9\,--\,11 years (Schwabe cycle), directions of the arrows are mainly toward the right, indicating that the solar-cycle length in the opposite hemispheres is unequal. However, this region is common to both the hemispheres and is strongly correlated, as seen by the WTC plot. A quasi-period in the range of  5\,--\,7 years is common and statistically significant between 1940 to 1985 with phase variations.

\subsection{Determination of the Gnevyshev Gap (GG)}

\begin{table}[!h]
	\caption{The detected Gnevyshev gap for the whole disk, the northern- and 
southern-hemisphere data. The $\times$ symbol represents the presence of the gap and ``--'' symbol represents that there is no Gnevyshev gap.}
	
	\label{tab:3}
	\begin{tabular}{llll}     
		
		\hline
		Solar Cycle No. &  Whole disk & N-hemisphere & S-hemisphere \\
		\hline
16 & $\times$ &  $\times$ &  $\times$ \\
17 & $\times$ &  --  &  $\times$  \\
18 & $\times$ &  $\times$ &  $\times$ \\
19 & --        &   $\times$ & -- \\
20 & --        &   $\times$ & -- \\
21 & $\times$ &  $\times$ &  $\times$ \\
22 & $\times$ &  $\times$ &  $\times$ \\
23 & $\times$ &  $\times$ &  $\times$ \\

\hline
\end{tabular}
\end{table}

Figure~10 shows the four month smoothed sunspot area plotted against time for Cycles 20 and 21. In this figure the existence of Gnevyshev gaps in the whole disk and also in both of the hemispheres data is clearly seen.  We have noticed that six out of eight solar cycles have shown a detectable Gnevyshev gap in the total sunspot-area time series. Seven (six) cycles have this gap in the data of northern (southern) hemisphere. We may conclude that this double-peaked nature of solar cycles, and the ensuing Gnevyshev gap, is not an artifact generated by averaging data from hemispheres that are slightly out of phase. Rather, we confirm earlier results \citep{2010SoPh..261..193N} that it is an inherent property in each hemisphere and must be caused by some physical mechanism occurring within each hemisphere.

\begin{figure}[!h]
\begin{center}
\includegraphics[width=0.9\textwidth]{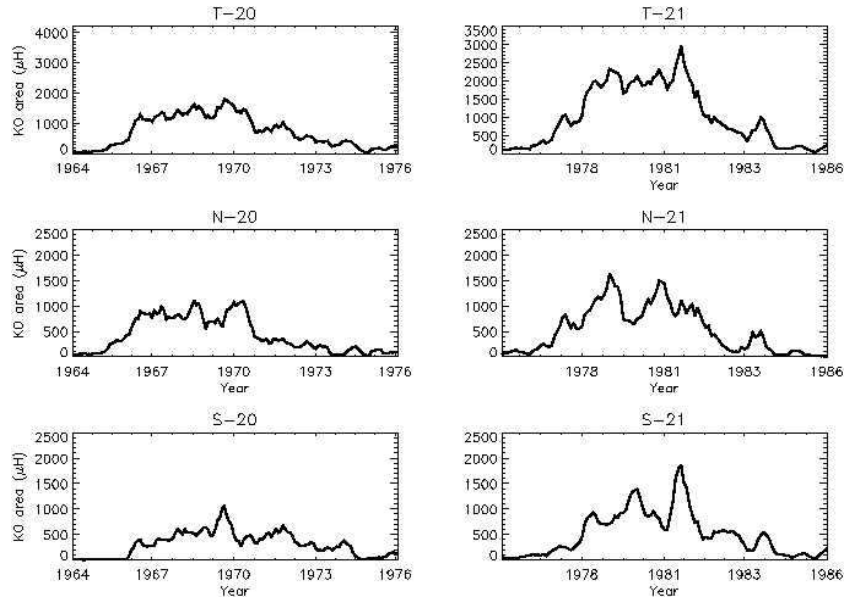} 
\caption{The four-month smoothed sunspot-area data are used to identify the Gnevyshev gaps and is shown here for the whole Sun (top), northern hemisphere (middle) and southern hemisphere (bottom) for Cycles 20 and 21.}
\end{center}
\label{fig:10}
\end{figure}

\section{Discussion and Conclusions}
Sunspot area is a key indicator of the solar-surface magnetic field and its variation \citep{2015LRSP...12....4H}. Long-term study of sunspot-area helps us better understand the solar-cycle evolution of the magnetic field, which is the main source of all types of activity in the Sun. We have utilized the newly digitized Kodaikanal Observatory sunspot-area data to investigate the characteristics of north--south asymmetry in solar activity during Solar Cycles 16\,--\,23, which may provide us valuable information about the nature of the solar dynamo between the opposite hemispheres in different solar cycles.
 So far, to the best of our knowledge, this is the first analysis of the asymmetric behavior, dominant hemisphere, phase asynchronicity, and the Gnevyshev Gap (GG) of this Kodaikanal data covering the eight Solar Cycles 16 to 23. Below, we have made a summary about the main findings with discussions:
i) It is observed that both the strength and temporal variation of the KOSO sunspot-area are not the same in the northern and southern hemispheres.  The peak time is also different in both hemispheres in many cycles. In both hemispheres the maximum and minimum strength of sunspot-areas were found during Cycles 19 and 20, respectively. The monthly mean sunspot-area favors the northern hemisphere during Solar Cycles 16, 19, and 20. However, 13-month average data indicated that both hemispheres had nearly equal strength in Cycles 17 and 18. 
ii)   N--S asymmetry is maximum around the minima or the onset of any new cycle, and it has a cyclic trend. The northern hemisphere was dominant during Cycle 19 and 20. The hemispheric asymmetry was not statistically significant in other cycles. Our cumulative analysis revealed that the northern hemisphere was more active in all cycles except in Cycle 20. 
iii)  Sunspot activity in the northern hemisphere leads in Solar Cycles 20, 21, and 22 by 12, 15, and 2 months, respectively. We did not find any time delay in Cycles 16, 17, 18, 19, and 23.
iv)  By using the wavelet technique, we have detected several midterm quasi-periodicities, along with the 11-year variations in the asymmetry. Our cross-wavelet analysis indicated that these periods are common and statistically significant in both the hemispheres with phase mixing. The presence of periods in the range of five years in both local wavelet and wavelet coherence plots may be assumed as a 
manifestation of the asymmetric nature of solar cycles or due to the double-peak structure of solar cycles \citep{2011ISRAA2011E...2G}. We found that phase coherency exists only in a narrow period belt of 9\,--\,12 years, including the 11-year Schwabe cycle.
v)  Finally, we have detected the Gnevyshev Gap in both of the hemispheres separately for individual Cycles 16 to 23. We have noticed that the Gnevyshev Gap is common in the 
sunspot-area data of the whole solar disk, northern, and in the southern hemisphere during Cycles 16, 18, 21, 22, and 23. This gap is absent during Cycle 17 in the northern hemisphere and during Cycle 19 and 20 for the southern hemisphere. Our results are consistent with the earlier detection of this phenomenon in the Greenwich sunspot-area data set \citep{2010SoPh..261..193N}. 

There is currently no clear physical explanation about the origin of different types of quasi-periods in various solar proxies. We have detected Rieger-type and QBOs in the present study, which has also been reported in the literature (\citealp{2002A&A...394..701K,  2005SoPh..226..359C, 2005A&A...438.1067K, 2010ApJ...718L..19F, 2011NewA...16..357B, 2012ApJ...749...27V,  2013ApJ...768..188C, 2014SoPh..289.4365K,   2016ApJ...830L..33M, 2017ApJ...845..137G, 2019SoPh..294..142C, 2020MNRAS.494.4930D} and references therein). Probably the midterm periods are manifested by emergence of magnetic flux at regular intervals from below the solar surface \citep{2002ApJ...566..505B} or by the excitation of Rossby-type waves, (e.g. \opencite{2003MNRAS.345..809L, 2005A&A...438.1067K, 2009MNRAS.392.1159C, 2010ApJ...724L..95Z, 2016ApJ...826...55G}). Recently, such Rossby-type waves in both solar upper and interior layers were detected and diagnosed \citep{2017NatAs...1E..86M, 2020A&A...635A.109H}. \cite{2015ApJ...805L..14Z} and \cite{2018ApJ...856...32Z} have suggested that slow magneto-Rossby waves could be responsible for long-term variations of solar activity. Thus, the periods observed in the hemispheric asymmetry might correspond to different harmonics of global slow magneto-Rossby waves. The QBOs were detected in the deep solar interior and on different solar surfaces and considered a global phenomenon (e.g. \opencite{2014SSRv..186..359B}).  However, the origin of QBOs due to the second dynamo's presence near the solar surface has also been suggested \citep{1998ApJ...509L..49B, 2014ARep...58..936O, 2016ApJ...826..138B, 2018ApJ...863...35S}. Recently, \cite{2019A&A...625A.117I} indicated that turbulent $\alpha$-dynamos with the Lorentz force are mainly responsible for generating the spatio-temporal behavior of the QBOs. However, to shed more light on the physical mechanisms behind these midterm quasi-periodicities, more investigation on solar interior dynamics is required.

We have found asynchronous behavior and time lead/lag in hemispheric solar activity. This phase shift is probably related to the hemispheric asymmetry with a connection to the variation of the about 11-year Schwabe cycle (\citealp{2006A&A...447..735T, 2006A&A...449L...1Z, 2014ApJ...796L..19M, 2018A&A...618A..89S} and references therein). However, our findings of cycle-wise dominant hemispheres and time delay between the hemispheres are not entirely in agreement with the previous results with ESAI sunspot-area data \citep{2015AJ....150...74Z} and sunspot numbers \citep{2019SoPh..294..142C}. But, our cumulative plots with KOSO data for different solar cycles are totally consistent with other recent long-term studies for 
sunspot-related activities measured at other observatories (e,g. \citealp{2019ApJ...875...90L, 2019SoPh..294..142C}, etc.). Moreover, our finding of phase coherency only in the small region of the Schwabe cycle period satisfies the conclusions of \cite{2007A&A...475L..33D}.

The reason behind the N--S asymmetry is still unclear. Our findings of time delay (phase difference) and different dominant hemispheres are probably caused by the unusual behavior of the dynamo mechanism in the opposite hemispheres. Attempts to explain the hemispheric asymmetry have been proposed in dynamo simulations in different ways such as asymmetric meridional circulation \citep{2010Sci...327.1350H, 2013ApJ...765..146M}; fluctuation and randomness in the formation of polar field \citep{2014SSRv..186..251N}; asymmetry in dipole and quadrupole components in solar polar-field reversal \citep{2017ApJ...835...84S}. Recently, \cite{2019MNRAS.489.4329H} explained the hemispheric asymmetry as a nonlinear coupling between the dipolar and quadrupolar components of the solar magnetic fields and noticed that global parity flips and hemispheric asymmetry is closely related. They particularly noted that if one hemisphere is stronger over another hemisphere for a sufficiently long time, parity change takes place. At the same time, an asymmetric variation of subsurface zonal flows in different latitude bands \citep{2014SoPh..289.3435K, 2014ApJ...789L...7Z, 2020SoPh..295...47K} might be related with this N--S asymmetry. However, these speculations are still far from determining the exact cause of the results obtained from our analysis. More observations and numerical simulations are required to understand this phase asynchrony better.

To conclude, we have analyzed about 100 years of digitized white-light data observed from Kodaikanal Observatory and extracted sunspot area to investigate the properties of the northern and southern hemispheres separately. We found some noticeable differences between the northern and southern hemisphere temporal behavior during the different phases of the solar cycle. In the future, we would like to investigate other long-term sunspot-related studies using Kodaiakanal white-light sunspot-area data and compare the results with other data sets, which may help us understand the underlying physical processes responsible for the observed long-term behaviour of sunspots. 

{\bf Acknowledgements}
{\small We thank the reviewer for fruitful comments.
We thank all of the observers at the Kodaikanal Solar Observatory who have observed continuously over a century. We also thank the people who have initiated the digitization program at Kodaikanal Observatory.}


%
%
\bibliographystyle{spr-mp-sola}

\begin{thebibliography}{}
\ifx\bisbn     \undefined \def\bisbn  #1{ISBN #1}\fi
\ifx\binits    \undefined \def\binits#1{#1}\fi
\ifx\bauthor   \undefined \def\bauthor#1{#1}\fi
\ifx\batitle   \undefined \def\batitle#1{#1}\fi
\ifx\bjtitle   \undefined \def\bjtitle#1{\textit{#1}}\fi
\ifx\bvolume   \undefined \def\bvolume#1{\textbf{#1}}\fi
\ifx\byear     \undefined \def\byear#1{#1}\fi
\ifx\bissue    \undefined \def\bissue#1{#1}\fi
\ifx\bfpage    \undefined \def\bfpage#1{#1}\fi
\ifx\blpage    \undefined \def\blpage #1{#1}\fi
\ifx\burl      \undefined \def\burl#1{#1}\fi
\ifx\href      \undefined \def\href#1#2{#2}\fi
\ifx\betal     \undefined \def\betal{et al.}\fi
\ifx\bctitle   \undefined \def\bctitle#1{#1}\fi
\ifx\beditor   \undefined \def\beditor#1{#1}\fi
\ifx\bbtitle   \undefined \def\bbtitle#1{\textit{#1}}\fi
\ifx\bedition  \undefined \def\bedition#1{#1}\fi
\ifx\bseriesno \undefined \def\bseriesno#1{\textbf{#1}}\fi
\ifx\blocation \undefined \def\blocation#1{#1}\fi
\ifx\bsertitle \undefined \def\bsertitle#1{\textit{#1}}\fi
\ifx\bsnm      \undefined \def\bsnm#1{#1}\fi
\ifx\bsuffix   \undefined \def\bsuffix#1{#1}\fi
\ifx\bparticle \undefined \def\bparticle#1{#1}\fi
\ifx\barticle  \undefined \def\barticle#1{}\fi
\ifx\binstitute  \undefined \def\binstitute#1{#1}\fi
\ifx\bpublisher  \undefined \def\bpublisher#1{#1}\fi
\ifx\doiurl    \undefined \def\doiurl#1{\href{#1}{DOI}}\fi
\makeatletter
\def\safeHref#1#2#3{\in@{http}{#2}\ifin@\href{#2}{#3}\else\href{#1#2}{#3}\fi}
\makeatother
\ifx\adsurl    \undefined
  \def\adsurl#1{\safeHref{https://ui.adsabs.harvard.edu/abs/}{#1}{ADS}}\fi
\ifx\arxivurl  \undefined
  \def\arxivurl#1{\safeHref{http://arxiv.org/abs/}{#1}{arXiv}}\fi
\ifx\botherref \undefined \def\botherref#1{}\fi
\ifx\url       \undefined \def\url#1{#1}\fi
\ifx\bchapter  \undefined \def\bchapter#1{}\fi
\ifx\bbook     \undefined \def\bbook#1{}\fi
\ifx\bcomment  \undefined \def\bcomment#1{#1}\fi
\ifx\oauthor   \undefined \def\oauthor#1{#1}\fi
\ifx\citeauthoryear \undefined\def \citeauthoryear#1{#1}\fi
\def\endbibitem {}
\ifx\bconflocation  \undefined \def\bconflocation#1{#1} \fi

\bibitem[\protect\citeauthoryear{{Ata{\c{c}}} and
  {{\"O}zg{\"u}{\c{c}}}}{1996}]{1996SoPh..166..201A}
\begin{barticle}
\bauthor{\bsnm{{Ata{\c{c}}}}, \binits{T.}},
\bauthor{\bsnm{{{\"O}zg{\"u}{\c{c}}}}, \binits{A.}}:
\byear{1996},
\batitle{{North-South Asymmetry in the Solar Flare Index}}.
\bjtitle{\solphys}
\bvolume{166},
\bfpage{201}.
\doiurl{https://doi.org/10.1007/BF00179363}.
\adsurl{1996SoPh..166..201A}.
\end{barticle}
\endbibitem

\bibitem[\protect\citeauthoryear{{Badalyan} and
  {Obridko}}{2011}]{2011NewA...16..357B}
\begin{barticle}
\bauthor{\bsnm{{Badalyan}}, \binits{O.G.}},
\bauthor{\bsnm{{Obridko}}, \binits{V.N.}}:
\byear{2011},
\batitle{{North-South asymmetry of the sunspot indices and its quasi-biennial
  oscillations}}.
\bjtitle{\na}
\bvolume{16},
\bfpage{357}.
\doiurl{https://doi.org/10.1016/j.newast.2011.01.005}.
\adsurl{2011NewA...16..357B}.
\end{barticle}
\endbibitem

\bibitem[\protect\citeauthoryear{{Badalyan} and
  {Obridko}}{2017}]{2017A&A...603A.109B}
\begin{barticle}
\bauthor{\bsnm{{Badalyan}}, \binits{O.G.}},
\bauthor{\bsnm{{Obridko}}, \binits{V.N.}}:
\byear{2017},
\batitle{{North-south asymmetry of solar activity as a superposition of two
  realizations - the sign and absolute value}}.
\bjtitle{\aap}
\bvolume{603},
\bfpage{A109}.
\doiurl{https://doi.org/10.1051/0004-6361/201527790}.
\adsurl{2017A&A...603A.109B}.
\end{barticle}
\endbibitem

\bibitem[\protect\citeauthoryear{{Ballester}, {Oliver}, and
  {Carbonell}}{2002}]{2002ApJ...566..505B}
\begin{barticle}
\bauthor{\bsnm{{Ballester}}, \binits{J.L.}},
\bauthor{\bsnm{{Oliver}}, \binits{R.}},
\bauthor{\bsnm{{Carbonell}}, \binits{M.}}:
\byear{2002},
\batitle{{The Near 160 Day Periodicity in the Photospheric Magnetic Flux}}.
\bjtitle{\apj}
\bvolume{566},
\bfpage{505}.
\doiurl{https://doi.org/10.1086/338075}.
\adsurl{2002ApJ...566..505B}.
\end{barticle}
\endbibitem

\bibitem[\protect\citeauthoryear{{Ballester}, {Oliver}, and
  {Carbonell}}{2005}]{2005A&A...431L...5B}
\begin{barticle}
\bauthor{\bsnm{{Ballester}}, \binits{J.L.}},
\bauthor{\bsnm{{Oliver}}, \binits{R.}},
\bauthor{\bsnm{{Carbonell}}, \binits{M.}}:
\byear{2005},
\batitle{{The periodic behaviour of the North-South asymmetry of sunspot areas
  revisited}}.
\bjtitle{\aap}
\bvolume{431},
\bfpage{L5}.
\doiurl{https://doi.org/10.1051/0004-6361:200400135}.
\adsurl{2005A&A...431L...5B}.
\end{barticle}
\endbibitem

\bibitem[\protect\citeauthoryear{{Baranyi}}{2015}]{2015MNRAS.447.1857B}
\begin{barticle}
\bauthor{\bsnm{{Baranyi}}, \binits{T.}}:
\byear{2015},
\batitle{{Comparison of Debrecen and Mount Wilson/Kodaikanal sunspot group tilt
  angles and the Joy's law}}.
\bjtitle{\mnras}
\bvolume{447},
\bfpage{1857}.
\doiurl{https://doi.org/10.1093/mnras/stu2572}.
\adsurl{2015MNRAS.447.1857B}.
\end{barticle}
\endbibitem

\bibitem[\protect\citeauthoryear{{Baranyi}, {Gy{\H{o}}ri}, and
  {Ludm{\'a}ny}}{2016}]{2016SoPh..291.3081B}
\begin{barticle}
\bauthor{\bsnm{{Baranyi}}, \binits{T.}},
\bauthor{\bsnm{{Gy{\H{o}}ri}}, \binits{L.}},
\bauthor{\bsnm{{Ludm{\'a}ny}}, \binits{A.}}:
\byear{2016},
\batitle{{On-line Tools for Solar Data Compiled at the Debrecen Observatory and
  Their Extensions with the Greenwich Sunspot Data}}.
\bjtitle{\solphys}
\bvolume{291},
\bfpage{3081}.
\doiurl{https://doi.org/10.1007/s11207-016-0930-1}.
\adsurl{2016SoPh..291.3081B}.
\end{barticle}
\endbibitem

\bibitem[\protect\citeauthoryear{{Bazilevskaya}
  et~al.}{2014}]{2014SSRv..186..359B}
\begin{barticle}
\bauthor{\bsnm{{Bazilevskaya}}, \binits{G.}},
\bauthor{\bsnm{{Broomhall}}, \binits{A.-M.}},
\bauthor{\bsnm{{Elsworth}}, \binits{Y.}},
\bauthor{\bsnm{{Nakariakov}}, \binits{V.M.}}:
\byear{2014},
\batitle{{A Combined Analysis of the Observational Aspects of the
  Quasi-biennial Oscillation in Solar Magnetic Activity}}.
\bjtitle{\ssr}
\bvolume{186},
\bfpage{359}.
\doiurl{https://doi.org/10.1007/s11214-014-0068-0}.
\adsurl{2014SSRv..186..359B}.
\end{barticle}
\endbibitem

\bibitem[\protect\citeauthoryear{{Beaudoin} et~al.}{2016}]{2016ApJ...826..138B}
\begin{barticle}
\bauthor{\bsnm{{Beaudoin}}, \binits{P.}},
\bauthor{\bsnm{{Simard}}, \binits{C.}},
\bauthor{\bsnm{{Cossette}}, \binits{J.-F.}},
\bauthor{\bsnm{{Charbonneau}}, \binits{P.}}:
\byear{2016},
\batitle{{Double Dynamo Signatures in a Global MHD Simulation and Mean-field
  Dynamos}}.
\bjtitle{\apj}
\bvolume{826},
\bfpage{138}.
\doiurl{https://doi.org/10.3847/0004-637X/826/2/138}.
\adsurl{2016ApJ...826..138B}.
\end{barticle}
\endbibitem

\bibitem[\protect\citeauthoryear{{Benevolenskaya}}{1998}]{1998ApJ...509L..49B}
\begin{barticle}
\bauthor{\bsnm{{Benevolenskaya}}, \binits{E.E.}}:
\byear{1998},
\batitle{{A Model of the Double Magnetic Cycle of the Sun}}.
\bjtitle{\apjl}
\bvolume{509},
\bfpage{L49}.
\doiurl{https://doi.org/10.1086/311755}.
\adsurl{1998ApJ...509L..49B}.
\end{barticle}
\endbibitem

\bibitem[\protect\citeauthoryear{{Cadavid} et~al.}{2005}]{2005SoPh..226..359C}
\begin{barticle}
\bauthor{\bsnm{{Cadavid}}, \binits{A.C.}},
\bauthor{\bsnm{{Lawrence}}, \binits{J.K.}},
\bauthor{\bsnm{{McDonald}}, \binits{D.P.}},
\bauthor{\bsnm{{Ruzmaikin}}, \binits{A.}}:
\byear{2005},
\batitle{{Independent Global Modes of Solar Magnetic Field Fluctuations}}.
\bjtitle{\solphys}
\bvolume{226},
\bfpage{359}.
\doiurl{https://doi.org/10.1007/s11207-005-8187-0}.
\adsurl{2005SoPh..226..359C}.
\end{barticle}
\endbibitem

\bibitem[\protect\citeauthoryear{{Carbonell}, {Oliver}, and
  {Ballester}}{1993}]{1993A&A...274..497C}
\begin{barticle}
\bauthor{\bsnm{{Carbonell}}, \binits{M.}},
\bauthor{\bsnm{{Oliver}}, \binits{R.}},
\bauthor{\bsnm{{Ballester}}, \binits{J.L.}}:
\byear{1993},
\batitle{{On the asymmetry of solar activity}}.
\bjtitle{\aap}
\bvolume{274},
\bfpage{497}.
\adsurl{1993A&A...274..497C}.
\end{barticle}
\endbibitem

\bibitem[\protect\citeauthoryear{{Carbonell}
  et~al.}{2007}]{2007A&A...476..951C}
\begin{barticle}
\bauthor{\bsnm{{Carbonell}}, \binits{M.}},
\bauthor{\bsnm{{Terradas}}, \binits{J.}},
\bauthor{\bsnm{{Oliver}}, \binits{R.}},
\bauthor{\bsnm{{Ballester}}, \binits{J.L.}}:
\byear{2007},
\batitle{{The statistical significance of the North-South asymmetry of solar
  activity revisited}}.
\bjtitle{\aap}
\bvolume{476},
\bfpage{951}.
\doiurl{https://doi.org/10.1051/0004-6361:20078004}.
\adsurl{2007A&A...476..951C}.
\end{barticle}
\endbibitem

\bibitem[\protect\citeauthoryear{{Chang} and {Glover}}{2010}]{2010NI..50..98C}
\begin{barticle}
\bauthor{\bsnm{{Chang}}, \binits{C.}},
\bauthor{\bsnm{{Glover}}, \binits{g.}}:
\byear{2010},
\batitle{{Time-frequency dynamics of resting-state brain connectivity measured
  with fMRI}}.
\bjtitle{\NI}
\bvolume{50},
\bfpage{98}.
\doiurl{https://doi.org/10.1016/j.neuroimage.2009.12.011}.
\end{barticle}
\endbibitem

\bibitem[\protect\citeauthoryear{{Charbonneau}}{2017}]{2017shin.confE.171C}
\begin{bchapter}
\bauthor{\bsnm{{Charbonneau}}, \binits{P.}}:
\byear{2017},
\bctitle{{Can the solar cycle be predicted?}}
In: \bbtitle{Solar Heliospheric and INterplanetary Environment (SHINE 2017)},
\bfpage{171}.
\adsurl{2017shin.confE.171C}.
\end{bchapter}
\endbibitem

\bibitem[\protect\citeauthoryear{{Choudhuri}}{1992}]{1992A&A...253..277C}
\begin{barticle}
\bauthor{\bsnm{{Choudhuri}}, \binits{A.R.}}:
\byear{1992},
\batitle{{Stochastic fluctuations of the solar dynamo}}.
\bjtitle{\aap}
\bvolume{253},
\bfpage{277}.
\adsurl{1992A&A...253..277C}.
\end{barticle}
\endbibitem

\bibitem[\protect\citeauthoryear{{Choudhuri}, {Chatterjee}, and
  {Jiang}}{2007}]{2007PhRvL..98m1103C}
\begin{barticle}
\bauthor{\bsnm{{Choudhuri}}, \binits{A.R.}},
\bauthor{\bsnm{{Chatterjee}}, \binits{P.}},
\bauthor{\bsnm{{Jiang}}, \binits{J.}}:
\byear{2007},
\batitle{{Predicting Solar Cycle 24 With a Solar Dynamo Model}}.
\bjtitle{\prl}
\bvolume{98},
\bfpage{131103}.
\doiurl{https://doi.org/10.1103/PhysRevLett.98.131103}.
\adsurl{2007PhRvL..98m1103C}.
\end{barticle}
\endbibitem

\bibitem[\protect\citeauthoryear{{Choudhuri}, {Schussler}, and
  {Dikpati}}{1995}]{1995A&A...303L..29C}
\begin{barticle}
\bauthor{\bsnm{{Choudhuri}}, \binits{A.R.}},
\bauthor{\bsnm{{Schussler}}, \binits{M.}},
\bauthor{\bsnm{{Dikpati}}, \binits{M.}}:
\byear{1995},
\batitle{{The solar dynamo with meridional circulation.}}
\bjtitle{\aap}
\bvolume{303},
\bfpage{L29}.
\adsurl{1995A&A...303L..29C}.
\end{barticle}
\endbibitem

\bibitem[\protect\citeauthoryear{{Chowdhury}, {Choudhary}, and
  {Gosain}}{2013}]{2013ApJ...768..188C}
\begin{barticle}
\bauthor{\bsnm{{Chowdhury}}, \binits{P.}},
\bauthor{\bsnm{{Choudhary}}, \binits{D.P.}},
\bauthor{\bsnm{{Gosain}}, \binits{S.}}:
\byear{2013},
\batitle{{A Study of the Hemispheric Asymmetry of Sunspot Area during Solar
  Cycles 23 and 24}}.
\bjtitle{\apj}
\bvolume{768},
\bfpage{188}.
\doiurl{https://doi.org/10.1088/0004-637X/768/2/188}.
\adsurl{2013ApJ...768..188C}.
\end{barticle}
\endbibitem

\bibitem[\protect\citeauthoryear{{Chowdhury}, {Khan}, and
  {Ray}}{2009}]{2009MNRAS.392.1159C}
\begin{barticle}
\bauthor{\bsnm{{Chowdhury}}, \binits{P.}},
\bauthor{\bsnm{{Khan}}, \binits{M.}},
\bauthor{\bsnm{{Ray}}, \binits{P.C.}}:
\byear{2009},
\batitle{{Intermediate-term periodicities in sunspot areas during solar cycles
  22 and 23}}.
\bjtitle{\mnras}
\bvolume{392},
\bfpage{1159}.
\doiurl{https://doi.org/10.1111/j.1365-2966.2008.14117.x}.
\adsurl{2009MNRAS.392.1159C}.
\end{barticle}
\endbibitem

\bibitem[\protect\citeauthoryear{{Chowdhury}
  et~al.}{2019}]{2019SoPh..294..142C}
\begin{barticle}
\bauthor{\bsnm{{Chowdhury}}, \binits{P.}},
\bauthor{\bsnm{{Kilcik}}, \binits{A.}},
\bauthor{\bsnm{{Yurchyshyn}}, \binits{V.}},
\bauthor{\bsnm{{Obridko}}, \binits{V.N.}},
\bauthor{\bsnm{{Rozelot}}, \binits{J.P.}}:
\byear{2019},
\batitle{{Analysis of the Hemispheric Sunspot Number Time Series for the Solar
  Cycles 18 to 24}}.
\bjtitle{\solphys}
\bvolume{294},
\bfpage{142}.
\doiurl{https://doi.org/10.1007/s11207-019-1530-7}.
\adsurl{2019SoPh..294..142C}.
\end{barticle}
\endbibitem

\bibitem[\protect\citeauthoryear{{Dasi-Espuig}
  et~al.}{2010}]{2010A&A...518A...7D}
\begin{barticle}
\bauthor{\bsnm{{Dasi-Espuig}}, \binits{M.}},
\bauthor{\bsnm{{Solanki}}, \binits{S.K.}},
\bauthor{\bsnm{{Krivova}}, \binits{N.A.}},
\bauthor{\bsnm{{Cameron}}, \binits{R.}},
\bauthor{\bsnm{{Pe{\~n}uela}}, \binits{T.}}:
\byear{2010},
\batitle{{Sunspot group tilt angles and the strength of the solar cycle}}.
\bjtitle{\aap}
\bvolume{518},
\bfpage{A7}.
\doiurl{https://doi.org/10.1051/0004-6361/201014301}.
\adsurl{2010A&A...518A...7D}.
\end{barticle}
\endbibitem

\bibitem[\protect\citeauthoryear{{Deng} et~al.}{2013a}]{2013Ap&SS.343...27D}
\begin{barticle}
\bauthor{\bsnm{{Deng}}, \binits{L.H.}},
\bauthor{\bsnm{{Gai}}, \binits{N.}},
\bauthor{\bsnm{{Tang}}, \binits{Y.K.}},
\bauthor{\bsnm{{Xu}}, \binits{C.L.}},
\bauthor{\bsnm{{Huang}}, \binits{W.J.}}:
\byear{2013}a,
\batitle{{Phase asynchrony of hemispheric flare activity revisited: empirical
  mode decomposition and wavelet transform analyses}}.
\bjtitle{\apss}
\bvolume{343},
\bfpage{27}.
\doiurl{https://doi.org/10.1007/s10509-012-1231-2}.
\adsurl{2013Ap&SS.343...27D}.
\end{barticle}
\endbibitem

\bibitem[\protect\citeauthoryear{{Deng} et~al.}{2013b}]{2013AN....334..217D}
\begin{barticle}
\bauthor{\bsnm{{Deng}}, \binits{L.H.}},
\bauthor{\bsnm{{Qu}}, \binits{Z.Q.}},
\bauthor{\bsnm{{Liu}}, \binits{T.}},
\bauthor{\bsnm{{Wang}}, \binits{K.R.}}:
\byear{2013}b,
\batitle{{The hemispheric variation of the flare index during solar cycles
  20-23}}.
\bjtitle{Astron. Nachr.}
\bvolume{334},
\bfpage{217}.
\doiurl{https://doi.org/10.1002/asna.201111786}.
\adsurl{2013AN....334..217D}.
\end{barticle}
\endbibitem

\bibitem[\protect\citeauthoryear{{Deng} et~al.}{2016}]{2016AJ....151...70D}
\begin{barticle}
\bauthor{\bsnm{{Deng}}, \binits{L.H.}},
\bauthor{\bsnm{{Xiang}}, \binits{Y.Y.}},
\bauthor{\bsnm{{Qu}}, \binits{Z.N.}},
\bauthor{\bsnm{{An}}, \binits{J.M.}}:
\byear{2016},
\batitle{{Systematic Regularity of Hemispheric Sunspot Areas Over the Past 140
  Years}}.
\bjtitle{\aj}
\bvolume{151},
\bfpage{70}.
\doiurl{https://doi.org/10.3847/0004-6256/151/3/70}.
\adsurl{2016AJ....151...70D}.
\end{barticle}
\endbibitem

\bibitem[\protect\citeauthoryear{{Deng} et~al.}{2019}]{2019MNRAS.488..111D}
\begin{barticle}
\bauthor{\bsnm{{Deng}}, \binits{L.H.}},
\bauthor{\bsnm{{Zhang}}, \binits{X.J.}},
\bauthor{\bsnm{{Li}}, \binits{G.Y.}},
\bauthor{\bsnm{{Deng}}, \binits{H.}},
\bauthor{\bsnm{{Wang}}, \binits{F.}}:
\byear{2019},
\batitle{{Phase and amplitude asymmetry in the quasi-biennial oscillation of
  solar H{\ensuremath{\alpha}} flare activity}}.
\bjtitle{\mnras}
\bvolume{488},
\bfpage{111}.
\doiurl{https://doi.org/10.1093/mnras/stz1653}.
\adsurl{2019MNRAS.488..111D}.
\end{barticle}
\endbibitem

\bibitem[\protect\citeauthoryear{{Deng} et~al.}{2020}]{2020MNRAS.494.4930D}
\begin{barticle}
\bauthor{\bsnm{{Deng}}, \binits{L.H.}},
\bauthor{\bsnm{{Fei}}, \binits{Y.}},
\bauthor{\bsnm{{Deng}}, \binits{H.}},
\bauthor{\bsnm{{Mei}}, \binits{Y.}},
\bauthor{\bsnm{{Wang}}, \binits{F.}}:
\byear{2020},
\batitle{{Spatial distribution of quasi-biennial oscillations in high-latitude
  solar activity}}.
\bjtitle{\mnras}
\bvolume{494},
\bfpage{4930}.
\doiurl{https://doi.org/10.1093/mnras/staa1061}.
\adsurl{2020MNRAS.494.4930D}.
\end{barticle}
\endbibitem

\bibitem[\protect\citeauthoryear{{Deng} et~al.}{2017}]{2017JSWSC...7A..34D}
\begin{barticle}
\bauthor{\bsnm{{Deng}}, \binits{L.}},
\bauthor{\bsnm{{Zhang}}, \binits{X.}},
\bauthor{\bsnm{{An}}, \binits{J.}},
\bauthor{\bsnm{{Cai}}, \binits{Y.}}:
\byear{2017},
\batitle{{Statistical properties of solar H{\ensuremath{\alpha}} flare
  activity}}.
\bjtitle{J. Space Weather Space Climate}
\bvolume{7},
\bfpage{A34}.
\doiurl{https://doi.org/10.1051/swsc/2017038}.
\adsurl{2017JSWSC...7A..34D}.
\end{barticle}
\endbibitem

\bibitem[\protect\citeauthoryear{{Dikpati} and
  {Charbonneau}}{1999}]{1999ApJ...518..508D}
\begin{barticle}
\bauthor{\bsnm{{Dikpati}}, \binits{M.}},
\bauthor{\bsnm{{Charbonneau}}, \binits{P.}}:
\byear{1999},
\batitle{{A Babcock-Leighton Flux Transport Dynamo with Solar-like Differential
  Rotation}}.
\bjtitle{\apj}
\bvolume{518},
\bfpage{508}.
\doiurl{https://doi.org/10.1086/307269}.
\adsurl{1999ApJ...518..508D}.
\end{barticle}
\endbibitem

\bibitem[\protect\citeauthoryear{{Donner} and
  {Thiel}}{2007}]{2007A&A...475L..33D}
\begin{barticle}
\bauthor{\bsnm{{Donner}}, \binits{R.}},
\bauthor{\bsnm{{Thiel}}, \binits{M.}}:
\byear{2007},
\batitle{{Scale-resolved phase coherence analysis of hemispheric sunspot
  activity: a new look at the north-south asymmetry}}.
\bjtitle{\aap}
\bvolume{475},
\bfpage{L33}.
\doiurl{https://doi.org/10.1051/0004-6361:20078672}.
\adsurl{2007A&A...475L..33D}.
\end{barticle}
\endbibitem

\bibitem[\protect\citeauthoryear{{Dorotovi{\v{c}}}
  et~al.}{2007}]{2007ASPC..368..527D}
\begin{bchapter}
\bauthor{\bsnm{{Dorotovi{\v{c}}}}, \binits{I.}},
\bauthor{\bsnm{{Journoud}}, \binits{P.}},
\bauthor{\bsnm{{Ryb{\'a}k}}, \binits{J.}},
\bauthor{\bsnm{{S{\'y}kora}}, \binits{J.}}:
\byear{2007},
\bctitle{{North-South Asymmetry of Ca II K Plages}}.
In: \beditor{\bsnm{{Heinzel}}, \binits{P.}},
\beditor{\bsnm{{Dorotovi{\v{c}}}}, \binits{I.}},
\beditor{\bsnm{{Rutten}}, \binits{R.J.}} (eds.)
\bbtitle{The Physics of Chromospheric Plasmas},
\bsertitle{Astron. Soc. Pacific, San Francisco}
\bseriesno{CS-368},
\bfpage{527}.
\adsurl{2007ASPC..368..527D}.
\end{bchapter}
\endbibitem

\bibitem[\protect\citeauthoryear{{Duchlev}}{2001}]{2001SoPh..199..211D}
\begin{barticle}
\bauthor{\bsnm{{Duchlev}}, \binits{P.I.}}:
\byear{2001},
\batitle{{An Estimation of the Long-Term Variation of a North-South Asymmetry
  of the Long-Lived Solar Filaments}}.
\bjtitle{\solphys}
\bvolume{199},
\bfpage{211}.
\doiurl{https://doi.org/10.1023/A:1010313817889}.
\adsurl{2001SoPh..199..211D}.
\end{barticle}
\endbibitem

\bibitem[\protect\citeauthoryear{{Ermolli} et~al.}{2014}]{2014SSRv..186..105E}
\begin{barticle}
\bauthor{\bsnm{{Ermolli}}, \binits{I.}},
\bauthor{\bsnm{{Shibasaki}}, \binits{K.}},
\bauthor{\bsnm{{Tlatov}}, \binits{A.}},
\bauthor{\bsnm{{van Driel-Gesztelyi}}, \binits{L.}}:
\byear{2014},
\batitle{{Solar Cycle Indices from the Photosphere to the Corona: Measurements
  and Underlying Physics}}.
\bjtitle{\ssr}
\bvolume{186},
\bfpage{105}.
\doiurl{https://doi.org/10.1007/s11214-014-0089-8}.
\adsurl{2014SSRv..186..105E}.
\end{barticle}
\endbibitem

\bibitem[\protect\citeauthoryear{{Farge}}{1992}]{1992ARFM...24...395F}
\begin{barticle}
\bauthor{\bsnm{{Farge}}, \binits{M.}}:
\byear{1992},
\batitle{{Wavelet Transforms and Their Applications to Turbulence}}.
\bjtitle{\arfm}
\bvolume{24},
\bfpage{90}.
\doiurl{https://doi.org/10.1146/annurev.fl.24.010192.002143}.
\adsurl{https://doi.org/10.1146/annurev.fl.24.010192.002143}.
\end{barticle}
\endbibitem

\bibitem[\protect\citeauthoryear{{Fletcher} et~al.}{2010}]{2010ApJ...718L..19F}
\begin{barticle}
\bauthor{\bsnm{{Fletcher}}, \binits{S.T.}},
\bauthor{\bsnm{{Broomhall}}, \binits{A.-M.}},
\bauthor{\bsnm{{Salabert}}, \binits{D.}},
\bauthor{\bsnm{{Basu}}, \binits{S.}},
\bauthor{\bsnm{{Chaplin}}, \binits{W.J.}},
\bauthor{\bsnm{{Elsworth}}, \binits{Y.}},
\bauthor{\bsnm{{Garcia}}, \binits{R.A.}},
\bauthor{\bsnm{{New}}, \binits{R.}}:
\byear{2010},
\batitle{{A Seismic Signature of a Second Dynamo?}}
\bjtitle{\apjl}
\bvolume{718},
\bfpage{L19}.
\doiurl{https://doi.org/10.1088/2041-8205/718/1/L19}.
\adsurl{2010ApJ...718L..19F}.
\end{barticle}
\endbibitem

\bibitem[\protect\citeauthoryear{{Gao}, {Li}, and
  {Shi}}{2009}]{2009MNRAS.400.1383G}
\begin{barticle}
\bauthor{\bsnm{{Gao}}, \binits{P.-X.}},
\bauthor{\bsnm{{Li}}, \binits{K.-J.}},
\bauthor{\bsnm{{Shi}}, \binits{X.-J.}}:
\byear{2009},
\batitle{{Hemispheric variation of coronal mass ejections in cycle 23}}.
\bjtitle{\mnras}
\bvolume{400},
\bfpage{1383}.
\doiurl{https://doi.org/10.1111/j.1365-2966.2009.15534.x}.
\adsurl{2009MNRAS.400.1383G}.
\end{barticle}
\endbibitem

\bibitem[\protect\citeauthoryear{{Georgieva}}{2011}]{2011ISRAA2011E...2G}
\begin{barticle}
\bauthor{\bsnm{{Georgieva}}, \binits{K.}}:
\byear{2011},
\batitle{{Why the Sunspot Cycle Is Double Peaked}}.
\bjtitle{ISRN Astron. Astrophy.}
\bvolume{2011},
\bfpage{437838}.
\adsurl{2011ISRAA2011E...2G}.
\end{barticle}
\endbibitem

\bibitem[\protect\citeauthoryear{{Gigolashvili}
  et~al.}{2005}]{2005SoPh..227...27G}
\begin{barticle}
\bauthor{\bsnm{{Gigolashvili}}, \binits{M.S.}},
\bauthor{\bsnm{{Japaridze}}, \binits{D.R.}},
\bauthor{\bsnm{{Mdzinarishvili}}, \binits{T.G.}},
\bauthor{\bsnm{{Chargeishvili}}, \binits{B.B.}}:
\byear{2005},
\batitle{{N S Asymmetry in the Solar Differential Rotation During 1957 1993}}.
\bjtitle{\solphys}
\bvolume{227},
\bfpage{27}.
\doiurl{https://doi.org/10.1007/s11207-005-1214-3}.
\adsurl{2005SoPh..227...27G}.
\end{barticle}
\endbibitem

\bibitem[\protect\citeauthoryear{{Gnevyshev}}{1963}]{1963SvA.....7..311G}
\begin{barticle}
\bauthor{\bsnm{{Gnevyshev}}, \binits{M.N.}}:
\byear{1963},
\batitle{{The Corona and the 11-Year Cycle of Solar Activity}}.
\bjtitle{\sovast}
\bvolume{7},
\bfpage{311}.
\adsurl{1963SvA.....7..311G}.
\end{barticle}
\endbibitem

\bibitem[\protect\citeauthoryear{{Gnevyshev}}{1967}]{1967SoPh....1..107G}
\begin{barticle}
\bauthor{\bsnm{{Gnevyshev}}, \binits{M.N.}}:
\byear{1967},
\batitle{{On the 11-Years Cycle of Solar Activity}}.
\bjtitle{\solphys}
\bvolume{1},
\bfpage{107}.
\doiurl{https://doi.org/10.1007/BF00150306}.
\adsurl{1967SoPh....1..107G}.
\end{barticle}
\endbibitem

\bibitem[\protect\citeauthoryear{{Gnevyshev} and
  {Ohl}}{1968}]{1968P&SS...16.1311G}
\begin{barticle}
\bauthor{\bsnm{{Gnevyshev}}, \binits{M.N.}},
\bauthor{\bsnm{{Ohl}}, \binits{A.I.}}:
\byear{1968},
\batitle{{On the manifestation of the 11-yr cycle of solar activity in the
  brightness of the nightglow}}.
\bjtitle{\planss}
\bvolume{16},
\bfpage{1311}.
\doiurl{https://doi.org/10.1016/0032-0633(68)90036-6}.
\adsurl{1968P&SS...16.1311G}.
\end{barticle}
\endbibitem

\bibitem[\protect\citeauthoryear{{Gon{\c{c}}alves}
  et~al.}{2014}]{2014SoPh..289.2283G}
\begin{barticle}
\bauthor{\bsnm{{Gon{\c{c}}alves}}, \binits{E.}},
\bauthor{\bsnm{{Mendes-Lopes}}, \binits{N.}},
\bauthor{\bsnm{{Dorotovi{\v{c}}}}, \binits{I.}},
\bauthor{\bsnm{{Fernandes}}, \binits{J.M.}},
\bauthor{\bsnm{{Garcia}}, \binits{A.}}:
\byear{2014},
\batitle{{North and South Hemispheric Solar Activity for Cycles 21 - 23:
  Asymmetry and Conditional Volatility of Plage Region Areas}}.
\bjtitle{\solphys}
\bvolume{289},
\bfpage{2283}.
\doiurl{https://doi.org/10.1007/s11207-013-0448-8}.
\adsurl{2014SoPh..289.2283G}.
\end{barticle}
\endbibitem

\bibitem[\protect\citeauthoryear{{Gray} et~al.}{2010}]{2010RvGeo..48.4001G}
\begin{barticle}
\bauthor{\bsnm{{Gray}}, \binits{L.J.}},
\bauthor{\bsnm{{Beer}}, \binits{J.}},
\bauthor{\bsnm{{Geller}}, \binits{M.}},
\bauthor{\bsnm{{Haigh}}, \binits{J.D.}},
\bauthor{\bsnm{{Lockwood}}, \binits{M.}},
\bauthor{\bsnm{{Matthes}}, \binits{K.}},
\bauthor{\bsnm{{Cubasch}}, \binits{U.}},
\bauthor{\bsnm{{Fleitmann}}, \binits{D.}},
\bauthor{\bsnm{{Harrison}}, \binits{G.}},
\bauthor{\bsnm{{Hood}}, \binits{L.}},
\bauthor{\bsnm{{Luterbacher}}, \binits{J.}},
\bauthor{\bsnm{{Meehl}}, \binits{G.A.}},
\bauthor{\bsnm{{Shindell}}, \binits{D.}},
\bauthor{\bsnm{{van Geel}}, \binits{B.}},
\bauthor{\bsnm{{White}}, \binits{W.}}:
\byear{2010},
\batitle{{Solar Influences on Climate}}.
\bjtitle{Rev. Geophys.}
\bvolume{48},
\bfpage{RG4001}.
\doiurl{https://doi.org/10.1029/2009RG000282}.
\adsurl{2010RvGeo..48.4001G}.
\end{barticle}
\endbibitem

\bibitem[\protect\citeauthoryear{{Grinsted}, {Moore}, and
  {Jevrejeva}}{2004}]{2004NPG..11..561G}
\begin{barticle}
\bauthor{\bsnm{{Grinsted}}, \binits{A.}},
\bauthor{\bsnm{{Moore}}, \binits{J.C.}},
\bauthor{\bsnm{{Jevrejeva}}, \binits{S.}}:
\byear{2004},
\batitle{{Applications of the Cross Wavelet Transform and Wavelet Coherence to
  Geophysical Time Series}}.
\bjtitle{\npg}
\bvolume{11},
\bfpage{6}.
\doiurl{https://doi.org/10.5194/npg-11-561-2004}.
\adsurl{https://npg.copernicus.org/articles/11/561/2004/}.
\end{barticle}
\endbibitem

\bibitem[\protect\citeauthoryear{{Gupta}, {Sivaraman}, and
  {Howard}}{1999}]{1999SoPh..188..225G}
\begin{barticle}
\bauthor{\bsnm{{Gupta}}, \binits{S.S.}},
\bauthor{\bsnm{{Sivaraman}}, \binits{K.R.}},
\bauthor{\bsnm{{Howard}}, \binits{R.F.}}:
\byear{1999},
\batitle{{Measurement of Kodaikanal White-Light Images - III. Rotation Rates
  and Activity Cycle Variations}}.
\bjtitle{\solphys}
\bvolume{188},
\bfpage{225}.
\doiurl{https://doi.org/10.1023/A:1005229124554}.
\adsurl{1999SoPh..188..225G}.
\end{barticle}
\endbibitem

\bibitem[\protect\citeauthoryear{{Gurgenashvili}
  et~al.}{2016}]{2016ApJ...826...55G}
\begin{barticle}
\bauthor{\bsnm{{Gurgenashvili}}, \binits{E.}},
\bauthor{\bsnm{{Zaqarashvili}}, \binits{T.V.}},
\bauthor{\bsnm{{Kukhianidze}}, \binits{V.}},
\bauthor{\bsnm{{Oliver}}, \binits{R.}},
\bauthor{\bsnm{{Ballester}}, \binits{J.L.}},
\bauthor{\bsnm{{Ramishvili}}, \binits{G.}},
\bauthor{\bsnm{{Shergelashvili}}, \binits{B.}},
\bauthor{\bsnm{{Hanslmeier}}, \binits{A.}},
\bauthor{\bsnm{{Poedts}}, \binits{S.}}:
\byear{2016},
\batitle{{Rieger-type Periodicity during Solar Cycles 14-24: Estimation of
  Dynamo Magnetic Field Strength in the Solar Interior}}.
\bjtitle{\apj}
\bvolume{826},
\bfpage{55}.
\doiurl{https://doi.org/10.3847/0004-637X/826/1/55}.
\adsurl{2016ApJ...826...55G}.
\end{barticle}
\endbibitem

\bibitem[\protect\citeauthoryear{{Gurgenashvili}
  et~al.}{2017}]{2017ApJ...845..137G}
\begin{barticle}
\bauthor{\bsnm{{Gurgenashvili}}, \binits{E.}},
\bauthor{\bsnm{{Zaqarashvili}}, \binits{T.V.}},
\bauthor{\bsnm{{Kukhianidze}}, \binits{V.}},
\bauthor{\bsnm{{Oliver}}, \binits{R.}},
\bauthor{\bsnm{{Ballester}}, \binits{J.L.}},
\bauthor{\bsnm{{Dikpati}}, \binits{M.}},
\bauthor{\bsnm{{McIntosh}}, \binits{S.W.}}:
\byear{2017},
\batitle{{North--South Asymmetry in Rieger-type Periodicity during Solar Cycles
  19 -- 23}}.
\bjtitle{\apj}
\bvolume{845},
\bfpage{137}.
\doiurl{https://doi.org/10.3847/1538-4357/aa830a}.
\adsurl{2017ApJ...845..137G}.
\end{barticle}
\endbibitem

\bibitem[\protect\citeauthoryear{{Hagino}, {Sakurai}, and
  {Miyazawa}}{2004}]{2004ASPC..325..157H}
\begin{bchapter}
\bauthor{\bsnm{{Hagino}}, \binits{M.}},
\bauthor{\bsnm{{Sakurai}}, \binits{T.}},
\bauthor{\bsnm{{Miyazawa}}, \binits{A.}}:
\byear{2004},
\bctitle{{Phase Relationship between the Activity Cycles of Sunspots and Polar
  Faculae}}.
In: \beditor{\bsnm{{Sakurai}}, \binits{T.}},
\beditor{\bsnm{{Sekii}}, \binits{T.}} (eds.)
\bbtitle{The Solar-B Mission and the Forefront of Solar Physics},
\bsertitle{Astron. Soc. Pacific, San Francisco}
\bseriesno{CS-325},
\bfpage{157}.
\adsurl{2004ASPC..325..157H}.
\end{bchapter}
\endbibitem

\bibitem[\protect\citeauthoryear{{Haigh}}{2007}]{2007LRSP....4....2H}
\begin{barticle}
\bauthor{\bsnm{{Haigh}}, \binits{J.D.}}:
\byear{2007},
\batitle{{The Sun and the Earth's Climate}}.
\bjtitle{Liv. Rev. Solar Phys.}
\bvolume{4},
\bfpage{2}.
\doiurl{https://doi.org/10.12942/lrsp-2007-2}.
\adsurl{2007LRSP....4....2H}.
\end{barticle}
\endbibitem

\bibitem[\protect\citeauthoryear{{Hanson}, {Gizon}, and
  {Liang}}{2020}]{2020A&A...635A.109H}
\begin{barticle}
\bauthor{\bsnm{{Hanson}}, \binits{C.S.}},
\bauthor{\bsnm{{Gizon}}, \binits{L.}},
\bauthor{\bsnm{{Liang}}, \binits{Z.-C.}}:
\byear{2020},
\batitle{{Solar Rossby waves observed in GONG++ ring-diagram flow maps}}.
\bjtitle{\aap}
\bvolume{635},
\bfpage{A109}.
\doiurl{https://doi.org/10.1051/0004-6361/201937321}.
\adsurl{2020A&A...635A.109H}.
\end{barticle}
\endbibitem

\bibitem[\protect\citeauthoryear{{Hao} et~al.}{2015}]{2015ApJS..221...33H}
\begin{barticle}
\bauthor{\bsnm{{Hao}}, \binits{Q.}},
\bauthor{\bsnm{{Fang}}, \binits{C.}},
\bauthor{\bsnm{{Cao}}, \binits{W.}},
\bauthor{\bsnm{{Chen}}, \binits{P.F.}}:
\byear{2015},
\batitle{{Statistical Analysis of Filament Features Based on the
  H{\ensuremath{\alpha}} Solar Images from 1988 to 2013 by Computer Automated
  Detection Method}}.
\bjtitle{\apjs}
\bvolume{221},
\bfpage{33}.
\doiurl{https://doi.org/10.1088/0067-0049/221/2/33}.
\adsurl{2015ApJS..221...33H}.
\end{barticle}
\endbibitem

\bibitem[\protect\citeauthoryear{{Hasan} et~al.}{2010}]{2010ASSP...19...12H}
\begin{barticle}
\bauthor{\bsnm{{Hasan}}, \binits{S.S.}},
\bauthor{\bsnm{{Mallik}}, \binits{D.C.V.}},
\bauthor{\bsnm{{Bagare}}, \binits{S.P.}},
\bauthor{\bsnm{{Rajaguru}}, \binits{S.P.}}:
\byear{2010},
\batitle{{Solar Physics at the Kodaikanal Observatory: A Historical
  Perspective}}.
\bjtitle{Astrophys. Space Sci. Proc.}
\bvolume{19},
\bfpage{12}.
\doiurl{https://doi.org/10.1007/978-3-642-02859-5_3}.
\adsurl{2010ASSP...19...12H}.
\end{barticle}
\endbibitem

\bibitem[\protect\citeauthoryear{{Hathaway}}{2015}]{2015LRSP...12....4H}
\begin{barticle}
\bauthor{\bsnm{{Hathaway}}, \binits{D.H.}}:
\byear{2015},
\batitle{{The Solar Cycle}}.
\bjtitle{Liv. Rev. Solar Phys.}
\bvolume{12},
\bfpage{4}.
\doiurl{https://doi.org/10.1007/lrsp-2015-4}.
\adsurl{2015LRSP...12....4H}.
\end{barticle}
\endbibitem

\bibitem[\protect\citeauthoryear{{Hathaway} and
  {Rightmire}}{2010}]{2010Sci...327.1350H}
\begin{barticle}
\bauthor{\bsnm{{Hathaway}}, \binits{D.H.}},
\bauthor{\bsnm{{Rightmire}}, \binits{L.}}:
\byear{2010},
\batitle{{Variations in the Sun{\textquoteright}s Meridional Flow over a Solar
  Cycle}}.
\bjtitle{Science}
\bvolume{327},
\bfpage{1350}.
\doiurl{https://doi.org/10.1126/science.1181990}.
\adsurl{2010Sci...327.1350H}.
\end{barticle}
\endbibitem

\bibitem[\protect\citeauthoryear{{Hazra} and
  {Nandy}}{2019}]{2019MNRAS.489.4329H}
\begin{barticle}
\bauthor{\bsnm{{Hazra}}, \binits{S.}},
\bauthor{\bsnm{{Nandy}}, \binits{D.}}:
\byear{2019},
\batitle{{The origin of parity changes in the solar cycle}}.
\bjtitle{\mnras}
\bvolume{489},
\bfpage{4329}.
\doiurl{https://doi.org/10.1093/mnras/stz2476}.
\adsurl{2019MNRAS.489.4329H}.
\end{barticle}
\endbibitem

\bibitem[\protect\citeauthoryear{{Hiremath} et~al.}{2020}]{2020ApJ...891..151H}
\begin{barticle}
\bauthor{\bsnm{{Hiremath}}, \binits{K.M.}},
\bauthor{\bsnm{{Rozelot}}, \binits{J.P.}},
\bauthor{\bsnm{{Sarp}}, \binits{V.}},
\bauthor{\bsnm{{Kilcik}}, \binits{A.}},
\bauthor{\bsnm{{Pavan}}, \binits{D.G.}},
\bauthor{\bsnm{{Gurumath}}, \binits{S.R.}}:
\byear{2020},
\batitle{{Nearly Century-scale Variation of the Sun's Radius}}.
\bjtitle{\apj}
\bvolume{891},
\bfpage{151}.
\doiurl{https://doi.org/10.3847/1538-4357/ab6d08}.
\adsurl{2020ApJ...891..151H}.
\end{barticle}
\endbibitem

\bibitem[\protect\citeauthoryear{{Howard}, {Gilman}, and
  {Gilman}}{1984}]{1984ApJ...283..373H}
\begin{barticle}
\bauthor{\bsnm{{Howard}}, \binits{R.}},
\bauthor{\bsnm{{Gilman}}, \binits{P.I.}},
\bauthor{\bsnm{{Gilman}}, \binits{P.A.}}:
\byear{1984},
\batitle{{Rotation of the sun measured from Mount Wilson white-light images}}.
\bjtitle{\apj}
\bvolume{283},
\bfpage{373}.
\doiurl{https://doi.org/10.1086/162315}.
\adsurl{1984ApJ...283..373H}.
\end{barticle}
\endbibitem

\bibitem[\protect\citeauthoryear{{Inceoglu} et~al.}{2019}]{2019A&A...625A.117I}
\begin{barticle}
\bauthor{\bsnm{{Inceoglu}}, \binits{F.}},
\bauthor{\bsnm{{Simoniello}}, \binits{R.}},
\bauthor{\bsnm{{Arlt}}, \binits{R.}},
\bauthor{\bsnm{{Rempel}}, \binits{M.}}:
\byear{2019},
\batitle{{Constraining non-linear dynamo models using quasi-biennial
  oscillations from sunspot area data}}.
\bjtitle{\aap}
\bvolume{625},
\bfpage{A117}.
\doiurl{https://doi.org/10.1051/0004-6361/201935272}.
\adsurl{2019A&A...625A.117I}.
\end{barticle}
\endbibitem

\bibitem[\protect\citeauthoryear{{Javaraiah}}{2007}]{2007MNRAS.377L..34J}
\begin{barticle}
\bauthor{\bsnm{{Javaraiah}}, \binits{J.}}:
\byear{2007},
\batitle{{North-south asymmetry in solar activity: predicting the amplitude of
  the next solar cycle}}.
\bjtitle{\mnras}
\bvolume{377},
\bfpage{L34}.
\doiurl{https://doi.org/10.1111/j.1745-3933.2007.00298.x}.
\adsurl{2007MNRAS.377L..34J}.
\end{barticle}
\endbibitem

\bibitem[\protect\citeauthoryear{{Javaraiah}}{2019}]{2019SoPh..294...64J}
\begin{barticle}
\bauthor{\bsnm{{Javaraiah}}, \binits{J.}}:
\byear{2019},
\batitle{{North-South Asymmetry in Solar Activity and Solar Cycle Prediction,
  IV: Prediction for Lengths of Upcoming Solar Cycles}}.
\bjtitle{\solphys}
\bvolume{294},
\bfpage{64}.
\doiurl{https://doi.org/10.1007/s11207-019-1442-6}.
\adsurl{2019SoPh..294...64J}.
\end{barticle}
\endbibitem

\bibitem[\protect\citeauthoryear{{Javaraiah}}{2020}]{2020SoPh..295....8J}
\begin{barticle}
\bauthor{\bsnm{{Javaraiah}}, \binits{J.}}:
\byear{2020},
\batitle{{Long-term Periodicities in North-south Asymmetry of Solar Activity
  and Alignments of the Giant Planets}}.
\bjtitle{\solphys}
\bvolume{295},
\bfpage{8}.
\doiurl{https://doi.org/10.1007/s11207-019-1575-7}.
\adsurl{2020SoPh..295....8J}.
\end{barticle}
\endbibitem

\bibitem[\protect\citeauthoryear{{Javaraiah} and
  {Gokhale}}{1997}]{1997SoPh..170..389J}
\begin{barticle}
\bauthor{\bsnm{{Javaraiah}}, \binits{J.}},
\bauthor{\bsnm{{Gokhale}}, \binits{M.H.}}:
\byear{1997},
\batitle{{Periodicities in the North-South Asymmetry of the Solar Differential
  Rotation and Surface Magnetic Field}}.
\bjtitle{\solphys}
\bvolume{170},
\bfpage{389}.
\doiurl{https://doi.org/10.1023/A:1004928020737}.
\adsurl{1997SoPh..170..389J}.
\end{barticle}
\endbibitem

\bibitem[\protect\citeauthoryear{{Jiang} et~al.}{2013}]{2013SSRv..176..289J}
\begin{barticle}
\bauthor{\bsnm{{Jiang}}, \binits{J.}},
\bauthor{\bsnm{{Cameron}}, \binits{R.H.}},
\bauthor{\bsnm{{Schmitt}}, \binits{D.}},
\bauthor{\bsnm{{Sch{\"u}ssler}}, \binits{M.}}:
\byear{2013},
\batitle{{Can Surface Flux Transport Account for the Weak Polar Field in Cycle
  23?}}
\bjtitle{\ssr}
\bvolume{176},
\bfpage{289}.
\doiurl{https://doi.org/10.1007/s11214-011-9783-y}.
\adsurl{2013SSRv..176..289J}.
\end{barticle}
\endbibitem

\bibitem[\protect\citeauthoryear{{Karak} and
  {Miesch}}{2017}]{2017ApJ...847...69K}
\begin{barticle}
\bauthor{\bsnm{{Karak}}, \binits{B.B.}},
\bauthor{\bsnm{{Miesch}}, \binits{M.}}:
\byear{2017},
\batitle{{Solar Cycle Variability Induced by Tilt Angle Scatter in a
  Babcock-Leighton Solar Dynamo Model}}.
\bjtitle{\apj}
\bvolume{847},
\bfpage{69}.
\doiurl{https://doi.org/10.3847/1538-4357/aa8636}.
\adsurl{2017ApJ...847...69K}.
\end{barticle}
\endbibitem

\bibitem[\protect\citeauthoryear{{Kilcik} et~al.}{2014}]{2014SoPh..289.4365K}
\begin{barticle}
\bauthor{\bsnm{{Kilcik}}, \binits{A.}},
\bauthor{\bsnm{{Ozguc}}, \binits{A.}},
\bauthor{\bsnm{{Yurchyshyn}}, \binits{V.}},
\bauthor{\bsnm{{Rozelot}}, \binits{J.P.}}:
\byear{2014},
\batitle{{Sunspot Count Periodicities in Different Zurich Sunspot Group Classes
  Since 1986}}.
\bjtitle{\solphys}
\bvolume{289},
\bfpage{4365}.
\doiurl{https://doi.org/10.1007/s11207-014-0580-0}.
\adsurl{2014SoPh..289.4365K}.
\end{barticle}
\endbibitem

\bibitem[\protect\citeauthoryear{{Knaack}, {Stenflo}, and
  {Berdyugina}}{2005}]{2005A&A...438.1067K}
\begin{barticle}
\bauthor{\bsnm{{Knaack}}, \binits{R.}},
\bauthor{\bsnm{{Stenflo}}, \binits{J.O.}},
\bauthor{\bsnm{{Berdyugina}}, \binits{S.V.}}:
\byear{2005},
\batitle{{Evolution and rotation of large-scale photospheric magnetic fields of
  the Sun during cycles 21-23. Periodicities, north-south asymmetries and
  r-mode signatures}}.
\bjtitle{\aap}
\bvolume{438},
\bfpage{1067}.
\doiurl{https://doi.org/10.1051/0004-6361:20042091}.
\adsurl{2005A&A...438.1067K}.
\end{barticle}
\endbibitem

\bibitem[\protect\citeauthoryear{{Komm}, {Howe}, and
  {Hill}}{2020}]{2020SoPh..295...47K}
\begin{barticle}
\bauthor{\bsnm{{Komm}}, \binits{R.}},
\bauthor{\bsnm{{Howe}}, \binits{R.}},
\bauthor{\bsnm{{Hill}}, \binits{F.}}:
\byear{2020},
\batitle{{Solar-Cycle Variation of the Subsurface Flows of Active- and
  Quiet-Region Subsets}}.
\bjtitle{\solphys}
\bvolume{295},
\bfpage{47}.
\doiurl{https://doi.org/10.1007/s11207-020-01611-5}.
\adsurl{2020SoPh..295...47K}.
\end{barticle}
\endbibitem

\bibitem[\protect\citeauthoryear{{Komm} et~al.}{2014}]{2014SoPh..289.3435K}
\begin{barticle}
\bauthor{\bsnm{{Komm}}, \binits{R.}},
\bauthor{\bsnm{{Howe}}, \binits{R.}},
\bauthor{\bsnm{{Gonz{\'a}lez Hern{\'a}ndez}}, \binits{I.}},
\bauthor{\bsnm{{Hill}}, \binits{F.}}:
\byear{2014},
\batitle{{Solar-Cycle Variation of Subsurface Zonal Flow}}.
\bjtitle{\solphys}
\bvolume{289},
\bfpage{3435}.
\doiurl{https://doi.org/10.1007/s11207-014-0490-1}.
\adsurl{2014SoPh..289.3435K}.
\end{barticle}
\endbibitem

\bibitem[\protect\citeauthoryear{{Krivova} and
  {Solanki}}{2002}]{2002A&A...394..701K}
\begin{barticle}
\bauthor{\bsnm{{Krivova}}, \binits{N.A.}},
\bauthor{\bsnm{{Solanki}}, \binits{S.K.}}:
\byear{2002},
\batitle{{The 1.3-year and 156-day periodicities in sunspot data: Wavelet
  analysis suggests a common origin}}.
\bjtitle{\aap}
\bvolume{394},
\bfpage{701}.
\doiurl{https://doi.org/10.1051/0004-6361:20021063}.
\adsurl{2002A&A...394..701K}.
\end{barticle}
\endbibitem

\bibitem[\protect\citeauthoryear{{Leussu} et~al.}{2017}]{2017A&A...599A.131L}
\begin{barticle}
\bauthor{\bsnm{{Leussu}}, \binits{R.}},
\bauthor{\bsnm{{Usoskin}}, \binits{I.G.}},
\bauthor{\bsnm{{Senthamizh Pavai}}, \binits{V.}},
\bauthor{\bsnm{{Diercke}}, \binits{A.}},
\bauthor{\bsnm{{Arlt}}, \binits{R.}},
\bauthor{\bsnm{{Denker}}, \binits{C.}},
\bauthor{\bsnm{{Mursula}}, \binits{K.}}:
\byear{2017},
\batitle{{Wings of the butterfly: Sunspot groups for 1826-2015}}.
\bjtitle{\aap}
\bvolume{599},
\bfpage{A131}.
\doiurl{https://doi.org/10.1051/0004-6361/201629533}.
\adsurl{2017A&A...599A.131L}.
\end{barticle}
\endbibitem

\bibitem[\protect\citeauthoryear{{Li}}{2009}]{2009SoPh..255..169L}
\begin{barticle}
\bauthor{\bsnm{{Li}}, \binits{K.J.}}:
\byear{2009},
\batitle{{Systematic Time Delay of Hemispheric Solar Activity}}.
\bjtitle{\solphys}
\bvolume{255},
\bfpage{169}.
\doiurl{https://doi.org/10.1007/s11207-009-9319-8}.
\adsurl{2009SoPh..255..169L}.
\end{barticle}
\endbibitem

\bibitem[\protect\citeauthoryear{{Li} et~al.}{2002}]{2002A&A...383..648L}
\begin{barticle}
\bauthor{\bsnm{{Li}}, \binits{K.J.}},
\bauthor{\bsnm{{Wang}}, \binits{J.X.}},
\bauthor{\bsnm{{Xiong}}, \binits{S.Y.}},
\bauthor{\bsnm{{Liang}}, \binits{H.F.}},
\bauthor{\bsnm{{Yun}}, \binits{H.S.}},
\bauthor{\bsnm{{Gu}}, \binits{X.M.}}:
\byear{2002},
\batitle{{Regularity of the north-south asymmetry of solar activity}}.
\bjtitle{\aap}
\bvolume{383},
\bfpage{648}.
\doiurl{https://doi.org/10.1051/0004-6361:20011799}.
\adsurl{2002A&A...383..648L}.
\end{barticle}
\endbibitem

\bibitem[\protect\citeauthoryear{{Li} et~al.}{2010}]{2010NewA...15..346L}
\begin{barticle}
\bauthor{\bsnm{{Li}}, \binits{K.J.}},
\bauthor{\bsnm{{Liu}}, \binits{X.H.}},
\bauthor{\bsnm{{Gao}}, \binits{P.X.}},
\bauthor{\bsnm{{Zhan}}, \binits{L.S.}}:
\byear{2010},
\batitle{{The north-south asymmetry of filaments in solar cycles 16-21}}.
\bjtitle{\na}
\bvolume{15},
\bfpage{346}.
\doiurl{https://doi.org/10.1016/j.newast.2009.09.009}.
\adsurl{2010NewA...15..346L}.
\end{barticle}
\endbibitem

\bibitem[\protect\citeauthoryear{{Li} et~al.}{2019}]{2019ApJ...875...90L}
\begin{barticle}
\bauthor{\bsnm{{Li}}, \binits{K.J.}},
\bauthor{\bsnm{{Xu}}, \binits{J.C.}},
\bauthor{\bsnm{{Yin}}, \binits{Z.Q.}},
\bauthor{\bsnm{{Feng}}, \binits{W.}}:
\byear{2019},
\batitle{{Why Does the Solar Corona Abnormally Rotate Faster Than the
  Photosphere?}}
\bjtitle{\apj}
\bvolume{875},
\bfpage{90}.
\doiurl{https://doi.org/10.3847/1538-4357/ab0f3a}.
\adsurl{2019ApJ...875...90L}.
\end{barticle}
\endbibitem

\bibitem[\protect\citeauthoryear{{Lou} et~al.}{2003}]{2003MNRAS.345..809L}
\begin{barticle}
\bauthor{\bsnm{{Lou}}, \binits{Y.-Q.}},
\bauthor{\bsnm{{Wang}}, \binits{Y.-M.}},
\bauthor{\bsnm{{Fan}}, \binits{Z.}},
\bauthor{\bsnm{{Wang}}, \binits{S.}},
\bauthor{\bsnm{{Wang}}, \binits{J.X.}}:
\byear{2003},
\batitle{{Periodicities in solar coronal mass ejections}}.
\bjtitle{\mnras}
\bvolume{345},
\bfpage{809}.
\doiurl{https://doi.org/10.1046/j.1365-8711.2003.06993.x}.
\adsurl{2003MNRAS.345..809L}.
\end{barticle}
\endbibitem

\bibitem[\protect\citeauthoryear{{Mandal} and
  {Banerjee}}{2016}]{2016ApJ...830L..33M}
\begin{barticle}
\bauthor{\bsnm{{Mandal}}, \binits{S.}},
\bauthor{\bsnm{{Banerjee}}, \binits{D.}}:
\byear{2016},
\batitle{{Sunspot Sizes and the Solar Cycle: Analysis Using Kodaikanal
  White-light Digitized Data}}.
\bjtitle{\apjl}
\bvolume{830},
\bfpage{L33}.
\doiurl{https://doi.org/10.3847/2041-8205/830/2/L33}.
\adsurl{2016ApJ...830L..33M}.
\end{barticle}
\endbibitem

\bibitem[\protect\citeauthoryear{{Mandal} et~al.}{2017}]{2017A&A...601A.106M}
\begin{barticle}
\bauthor{\bsnm{{Mandal}}, \binits{S.}},
\bauthor{\bsnm{{Hegde}}, \binits{M.}},
\bauthor{\bsnm{{Samanta}}, \binits{T.}},
\bauthor{\bsnm{{Hazra}}, \binits{G.}},
\bauthor{\bsnm{{Banerjee}}, \binits{D.}},
\bauthor{\bsnm{{Ravindra}}, \binits{B.}}:
\byear{2017},
\batitle{{Kodaikanal digitized white-light data archive (1921-2011): Analysis
  of various solar cycle features}}.
\bjtitle{\aap}
\bvolume{601},
\bfpage{A106}.
\doiurl{https://doi.org/10.1051/0004-6361/201628651}.
\adsurl{2017A&A...601A.106M}.
\end{barticle}
\endbibitem

\bibitem[\protect\citeauthoryear{{Maraun} and
  {Kurths}}{2004}]{2004NPG..11..505M}
\begin{barticle}
\bauthor{\bsnm{{Maraun}}, \binits{D.}},
\bauthor{\bsnm{{Kurths}}, \binits{D.}}:
\byear{2004},
\batitle{{Cross Wavelet Analysis: Significance Testing and Pitfalls}}.
\bjtitle{\npg}
\bvolume{11},
\bfpage{10}.
\doiurl{https://doi.org/10.5194/npg-11-505-2004}.
\adsurl{https://doi.org/10.5194/npg-11-505-2004}.
\end{barticle}
\endbibitem

\bibitem[\protect\citeauthoryear{{McClintock} and
  {Norton}}{2013}]{2013SoPh..287..215M}
\begin{barticle}
\bauthor{\bsnm{{McClintock}}, \binits{B.H.}},
\bauthor{\bsnm{{Norton}}, \binits{A.A.}}:
\byear{2013},
\batitle{{Recovering Joy's Law as a Function of Solar Cycle, Hemisphere, and
  Longitude}}.
\bjtitle{\solphys}
\bvolume{287},
\bfpage{215}.
\doiurl{https://doi.org/10.1007/s11207-013-0338-0}.
\adsurl{2013SoPh..287..215M}.
\end{barticle}
\endbibitem

\bibitem[\protect\citeauthoryear{{McIntosh} and
  {Leamon}}{2014}]{2014ApJ...796L..19M}
\begin{barticle}
\bauthor{\bsnm{{McIntosh}}, \binits{S.W.}},
\bauthor{\bsnm{{Leamon}}, \binits{R.J.}}:
\byear{2014},
\batitle{{On Magnetic Activity Band Overlap, Interaction, and the Formation of
  Complex Solar Active Regions}}.
\bjtitle{\apjl}
\bvolume{796},
\bfpage{L19}.
\doiurl{https://doi.org/10.1088/2041-8205/796/1/L19}.
\adsurl{2014ApJ...796L..19M}.
\end{barticle}
\endbibitem

\bibitem[\protect\citeauthoryear{{McIntosh} et~al.}{2013}]{2013ApJ...765..146M}
\begin{barticle}
\bauthor{\bsnm{{McIntosh}}, \binits{S.W.}},
\bauthor{\bsnm{{Leamon}}, \binits{R.J.}},
\bauthor{\bsnm{{Gurman}}, \binits{J.B.}},
\bauthor{\bsnm{{Olive}}, \binits{J.-P.}},
\bauthor{\bsnm{{Cirtain}}, \binits{J.W.}},
\bauthor{\bsnm{{Hathaway}}, \binits{D.H.}},
\bauthor{\bsnm{{Burkepile}}, \binits{J.}},
\bauthor{\bsnm{{Miesch}}, \binits{M.}},
\bauthor{\bsnm{{Markel}}, \binits{R.S.}},
\bauthor{\bsnm{{Sitongia}}, \binits{L.}}:
\byear{2013},
\batitle{{Hemispheric Asymmetries of Solar Photospheric Magnetism: Radiative,
  Particulate, and Heliospheric Impacts}}.
\bjtitle{\apj}
\bvolume{765},
\bfpage{146}.
\doiurl{https://doi.org/10.1088/0004-637X/765/2/146}.
\adsurl{2013ApJ...765..146M}.
\end{barticle}
\endbibitem

\bibitem[\protect\citeauthoryear{{McIntosh} et~al.}{2017}]{2017NatAs...1E..86M}
\begin{barticle}
\bauthor{\bsnm{{McIntosh}}, \binits{S.W.}},
\bauthor{\bsnm{{Cramer}}, \binits{W.J.}},
\bauthor{\bsnm{{Pichardo Marcano}}, \binits{M.}},
\bauthor{\bsnm{{Leamon}}, \binits{R.J.}}:
\byear{2017},
\batitle{{The detection of Rossby-like waves on the Sun}}.
\bjtitle{Nature Astron.}
\bvolume{1},
\bfpage{0086}.
\doiurl{https://doi.org/10.1038/s41550-017-0086}.
\adsurl{2017NatAs...1E..86M}.
\end{barticle}
\endbibitem

\bibitem[\protect\citeauthoryear{{Mordvinov}}{2007}]{2007SoPh..246..445M}
\begin{barticle}
\bauthor{\bsnm{{Mordvinov}}, \binits{A.V.}}:
\byear{2007},
\batitle{{Magnetic Flux Imbalance of the Solar and Heliospheric Magnetic
  Fields}}.
\bjtitle{\solphys}
\bvolume{246},
\bfpage{445}.
\doiurl{https://doi.org/10.1007/s11207-007-9082-7}.
\adsurl{2007SoPh..246..445M}.
\end{barticle}
\endbibitem

\bibitem[\protect\citeauthoryear{{Morlet} et~al.}{1982}]{1982GP...47...203M}
\begin{barticle}
\bauthor{\bsnm{{Morlet}}, \binits{J.}},
\bauthor{\bsnm{{Arens}}, \binits{G.}},
\bauthor{\bsnm{{Fourgeau}}},
\bauthor{\bsnm{{Giard}}, \binits{D.}}:
\byear{1982},
\batitle{{Wave propagation and sampling theory-Part I: Complex signal and
  scattering in multilayered media}}.
\bjtitle{\gp}
\bvolume{47},
\bfpage{203}.
\doiurl{https://doi.org/10.1190/1.1441329}.
\adsurl{https://doi.org/10.1190/1.1441329}.
\end{barticle}
\endbibitem

\bibitem[\protect\citeauthoryear{{Murak{\"o}zy} and
  {Ludm{\'a}ny}}{2012}]{2012MNRAS.419.3624M}
\begin{barticle}
\bauthor{\bsnm{{Murak{\"o}zy}}, \binits{J.}},
\bauthor{\bsnm{{Ludm{\'a}ny}}, \binits{A.}}:
\byear{2012},
\batitle{{Phase lags of solar hemispheric cycles}}.
\bjtitle{\mnras}
\bvolume{419},
\bfpage{3624}.
\doiurl{https://doi.org/10.1111/j.1365-2966.2011.20011.x}.
\adsurl{2012MNRAS.419.3624M}.
\end{barticle}
\endbibitem

\bibitem[\protect\citeauthoryear{{Nistic{\`o}}
  et~al.}{2015}]{2015A&A...583A.127N}
\begin{barticle}
\bauthor{\bsnm{{Nistic{\`o}}}, \binits{G.}},
\bauthor{\bsnm{{Zimbardo}}, \binits{G.}},
\bauthor{\bsnm{{Patsourakos}}, \binits{S.}},
\bauthor{\bsnm{{Bothmer}}, \binits{V.}},
\bauthor{\bsnm{{Nakariakov}}, \binits{V.M.}}:
\byear{2015},
\batitle{{North-south asymmetry in the magnetic deflection of polar coronal
  hole jets}}.
\bjtitle{\aap}
\bvolume{583},
\bfpage{A127}.
\doiurl{https://doi.org/10.1051/0004-6361/201525731}.
\adsurl{2015A&A...583A.127N}.
\end{barticle}
\endbibitem

\bibitem[\protect\citeauthoryear{{Norton} and
  {Gallagher}}{2010}]{2010SoPh..261..193N}
\begin{barticle}
\bauthor{\bsnm{{Norton}}, \binits{A.A.}},
\bauthor{\bsnm{{Gallagher}}, \binits{J.C.}}:
\byear{2010},
\batitle{{Solar-Cycle Characteristics Examined in Separate Hemispheres: Phase,
  Gnevyshev Gap, and Length of Minimum}}.
\bjtitle{\solphys}
\bvolume{261},
\bfpage{193}.
\doiurl{https://doi.org/10.1007/s11207-009-9479-6}.
\adsurl{2010SoPh..261..193N}.
\end{barticle}
\endbibitem

\bibitem[\protect\citeauthoryear{{Norton}, {Charbonneau}, and
  {Passos}}{2014}]{2014SSRv..186..251N}
\begin{barticle}
\bauthor{\bsnm{{Norton}}, \binits{A.A.}},
\bauthor{\bsnm{{Charbonneau}}, \binits{P.}},
\bauthor{\bsnm{{Passos}}, \binits{D.}}:
\byear{2014},
\batitle{{Hemispheric Coupling: Comparing Dynamo Simulations and
  Observations}}.
\bjtitle{\ssr}
\bvolume{186},
\bfpage{251}.
\doiurl{https://doi.org/10.1007/s11214-014-0100-4}.
\adsurl{2014SSRv..186..251N}.
\end{barticle}
\endbibitem

\bibitem[\protect\citeauthoryear{{Obridko} and
  {Badalyan}}{2014}]{2014ARep...58..936O}
\begin{barticle}
\bauthor{\bsnm{{Obridko}}, \binits{V.N.}},
\bauthor{\bsnm{{Badalyan}}, \binits{O.G.}}:
\byear{2014},
\batitle{{Cyclic and secular variations sunspot groups with various scales}}.
\bjtitle{Astronomy Reports}
\bvolume{58},
\bfpage{936}.
\doiurl{https://doi.org/10.1134/S1063772914120075}.
\adsurl{2014ARep...58..936O}.
\end{barticle}
\endbibitem

\bibitem[\protect\citeauthoryear{{Oloketuyi}, {Liu}, and
  {Zhao}}{2019}]{2019ApJ...874...20O}
\begin{barticle}
\bauthor{\bsnm{{Oloketuyi}}, \binits{J.}},
\bauthor{\bsnm{{Liu}}, \binits{Y.}},
\bauthor{\bsnm{{Zhao}}, \binits{M.}}:
\byear{2019},
\batitle{{The Periodic and Temporal Behaviors of Solar X-Ray Flares in Solar
  Cycles 23 and 24}}.
\bjtitle{\apj}
\bvolume{874},
\bfpage{20}.
\doiurl{https://doi.org/10.3847/1538-4357/ab064c}.
\adsurl{2019ApJ...874...20O}.
\end{barticle}
\endbibitem

\bibitem[\protect\citeauthoryear{{Passos} et~al.}{2014}]{2014A&A...563A..18P}
\begin{barticle}
\bauthor{\bsnm{{Passos}}, \binits{D.}},
\bauthor{\bsnm{{Nandy}}, \binits{D.}},
\bauthor{\bsnm{{Hazra}}, \binits{S.}},
\bauthor{\bsnm{{Lopes}}, \binits{I.}}:
\byear{2014},
\batitle{{A solar dynamo model driven by mean-field alpha and Babcock-Leighton
  sources: fluctuations, grand-minima-maxima, and hemispheric asymmetry in
  sunspot cycles}}.
\bjtitle{\aap}
\bvolume{563},
\bfpage{A18}.
\doiurl{https://doi.org/10.1051/0004-6361/201322635}.
\adsurl{2014A&A...563A..18P}.
\end{barticle}
\endbibitem

\bibitem[\protect\citeauthoryear{{Ravindra} and
  {Javaraiah}}{2015}]{2015NewA...39...55R}
\begin{barticle}
\bauthor{\bsnm{{Ravindra}}, \binits{B.}},
\bauthor{\bsnm{{Javaraiah}}, \binits{J.}}:
\byear{2015},
\batitle{{Hemispheric asymmetry of sunspot area in solar cycle 23 and rising
  phase of solar cycle 24: Comparison of three data sets}}.
\bjtitle{\na}
\bvolume{39},
\bfpage{55}.
\doiurl{https://doi.org/10.1016/j.newast.2015.03.004}.
\adsurl{2015NewA...39...55R}.
\end{barticle}
\endbibitem

\bibitem[\protect\citeauthoryear{{Ravindra} et~al.}{2013}]{2013A&A...550A..19R}
\begin{barticle}
\bauthor{\bsnm{{Ravindra}}, \binits{B.}},
\bauthor{\bsnm{{Priya}}, \binits{T.G.}},
\bauthor{\bsnm{{Amareswari}}, \binits{K.}},
\bauthor{\bsnm{{Priyal}}, \binits{M.}},
\bauthor{\bsnm{{Nazia}}, \binits{A.A.}},
\bauthor{\bsnm{{Banerjee}}, \binits{D.}}:
\byear{2013},
\batitle{{Digitized archive of the Kodaikanal images: Representative results of
  solar cycle variation from sunspot area determination}}.
\bjtitle{\aap}
\bvolume{550},
\bfpage{A19}.
\doiurl{https://doi.org/10.1051/0004-6361/201220416}.
\adsurl{2013A&A...550A..19R}.
\end{barticle}
\endbibitem

\bibitem[\protect\citeauthoryear{{Ravindra} et~al.}{2020}]{2020Ap&SS.365...14R}
\begin{barticle}
\bauthor{\bsnm{{Ravindra}}, \binits{B.}},
\bauthor{\bsnm{{Pichamani}}, \binits{K.}},
\bauthor{\bsnm{{Selvendran}}, \binits{R.}},
\bauthor{\bsnm{{Samuel}}, \binits{J.}},
\bauthor{\bsnm{{Kumar}}, \binits{P.}},
\bauthor{\bsnm{{Jassoria}}, \binits{N.}},
\bauthor{\bsnm{{Navneeth}}, \binits{R.S.}}:
\byear{2020},
\batitle{{Sunspot drawings at Kodaikanal Observatory: a representative results
  on hemispheric sunspot numbers and area measurements}}.
\bjtitle{\apss}
\bvolume{365},
\bfpage{14}.
\doiurl{https://doi.org/10.1007/s10509-020-3725-7}.
\adsurl{2020Ap&SS.365...14R}.
\end{barticle}
\endbibitem

\bibitem[\protect\citeauthoryear{{Richardson}, {von Rosenvinge}, and
  {Cane}}{2016}]{2016SoPh..291.2117R}
\begin{barticle}
\bauthor{\bsnm{{Richardson}}, \binits{I.G.}},
\bauthor{\bsnm{{von Rosenvinge}}, \binits{T.T.}},
\bauthor{\bsnm{{Cane}}, \binits{H.V.}}:
\byear{2016},
\batitle{{North/South Hemispheric Periodicities in the 25 MeV Solar Proton
  Event Rate During the Rising and Peak Phases of Solar Cycle 24}}.
\bjtitle{\solphys}
\bvolume{291},
\bfpage{2117}.
\doiurl{https://doi.org/10.1007/s11207-016-0948-4}.
\adsurl{2016SoPh..291.2117R}.
\end{barticle}
\endbibitem

\bibitem[\protect\citeauthoryear{{Roy}}{1977}]{1977SoPh...52...53R}
\begin{barticle}
\bauthor{\bsnm{{Roy}}, \binits{J.-R.}}:
\byear{1977},
\batitle{{The north-south distribution of major solar flare events, sunspot
  magnetic classes and sunspot areas (1955 1974)}}.
\bjtitle{\solphys}
\bvolume{52},
\bfpage{53}.
\doiurl{https://doi.org/10.1007/BF00935789}.
\adsurl{1977SoPh...52...53R}.
\end{barticle}
\endbibitem

\bibitem[\protect\citeauthoryear{{Scherrer} et~al.}{1995}]{1995SoPh..162..129S}
\begin{barticle}
\bauthor{\bsnm{{Scherrer}}, \binits{P.H.}},
\bauthor{\bsnm{{Bogart}}, \binits{R.S.}},
\bauthor{\bsnm{{Bush}}, \binits{R.I.}},
\bauthor{\bsnm{{Hoeksema}}, \binits{J.T.}},
\bauthor{\bsnm{{Kosovichev}}, \binits{A.G.}},
\bauthor{\bsnm{{Schou}}, \binits{J.}},
\bauthor{\bsnm{{Rosenberg}}, \binits{W.}},
\bauthor{\bsnm{{Springer}}, \binits{L.}},
\bauthor{\bsnm{{Tarbell}}, \binits{T.D.}},
\bauthor{\bsnm{{Title}}, \binits{A.}},
\bauthor{\bsnm{{Wolfson}}, \binits{C.J.}},
\bauthor{\bsnm{{Zayer}}, \binits{I.}},
\bauthor{\bsnm{{MDI Engineering Team}}}:
\byear{1995},
\batitle{{The Solar Oscillations Investigation - Michelson Doppler Imager}}.
\bjtitle{\solphys}
\bvolume{162},
\bfpage{129}.
\doiurl{https://doi.org/10.1007/BF00733429}.
\adsurl{1995SoPh..162..129S}.
\end{barticle}
\endbibitem

\bibitem[\protect\citeauthoryear{{Sch{\"u}ssler} and
  {Cameron}}{2018}]{2018A&A...618A..89S}
\begin{barticle}
\bauthor{\bsnm{{Sch{\"u}ssler}}, \binits{M.}},
\bauthor{\bsnm{{Cameron}}, \binits{R.H.}}:
\byear{2018},
\batitle{{Origin of the hemispheric asymmetry of solar activity}}.
\bjtitle{\aap}
\bvolume{618},
\bfpage{A89}.
\doiurl{https://doi.org/10.1051/0004-6361/201833532}.
\adsurl{2018A&A...618A..89S}.
\end{barticle}
\endbibitem

\bibitem[\protect\citeauthoryear{{Segu{\'\i}}
  et~al.}{2019}]{2019AdSpR..63.3738S}
\begin{barticle}
\bauthor{\bsnm{{Segu{\'\i}}}, \binits{A.}},
\bauthor{\bsnm{{Curto}}, \binits{J.J.}},
\bauthor{\bsnm{{de Paula}}, \binits{V.}},
\bauthor{\bsnm{{Rodr{\'\i}guez-Gas{\'e}n}}, \binits{R.}},
\bauthor{\bsnm{{Vaquero}}, \binits{J.M.}}:
\byear{2019},
\batitle{{Temporal variation and asymmetry of sunspot and solar plage types
  from 1930 to 1936}}.
\bjtitle{Adv. Space Res.}
\bvolume{63},
\bfpage{3738}.
\doiurl{https://doi.org/10.1016/j.asr.2019.02.018}.
\adsurl{2019AdSpR..63.3738S}.
\end{barticle}
\endbibitem

\bibitem[\protect\citeauthoryear{{Shukuya} and
  {Kusano}}{2017}]{2017ApJ...835...84S}
\begin{barticle}
\bauthor{\bsnm{{Shukuya}}, \binits{D.}},
\bauthor{\bsnm{{Kusano}}, \binits{K.}}:
\byear{2017},
\batitle{{Simulation Study of Hemispheric Phase-Asymmetry in the Solar Cycle}}.
\bjtitle{\apj}
\bvolume{835},
\bfpage{84}.
\doiurl{https://doi.org/10.3847/1538-4357/835/1/84}.
\adsurl{2017ApJ...835...84S}.
\end{barticle}
\endbibitem

\bibitem[\protect\citeauthoryear{{Sivaraman}, {Gupta}, and
  {Howard}}{1993}]{1993SoPh..146...27S}
\begin{barticle}
\bauthor{\bsnm{{Sivaraman}}, \binits{K.R.}},
\bauthor{\bsnm{{Gupta}}, \binits{S.S.}},
\bauthor{\bsnm{{Howard}}, \binits{R.F.}}:
\byear{1993},
\batitle{{Measurement of Kodiakanal White-Light Images - Part One}}.
\bjtitle{\solphys}
\bvolume{146},
\bfpage{27}.
\doiurl{https://doi.org/10.1007/BF00662168}.
\adsurl{1993SoPh..146...27S}.
\end{barticle}
\endbibitem

\bibitem[\protect\citeauthoryear{{Sivaraman}
  et~al.}{2003}]{2003SoPh..214...65S}
\begin{barticle}
\bauthor{\bsnm{{Sivaraman}}, \binits{K.R.}},
\bauthor{\bsnm{{Sivaraman}}, \binits{H.}},
\bauthor{\bsnm{{Gupta}}, \binits{S.S.}},
\bauthor{\bsnm{{Howard}}, \binits{R.F.}}:
\byear{2003},
\batitle{{Measurement of Kodaikanal white-light images - VI. Variation of
  Rotation Rate with Age of Sunspot Groups}}.
\bjtitle{\solphys}
\bvolume{214},
\bfpage{65}.
\doiurl{https://doi.org/10.1023/A:1024075100667}.
\adsurl{2003SoPh..214...65S}.
\end{barticle}
\endbibitem

\bibitem[\protect\citeauthoryear{{Sivaraman}
  et~al.}{2010}]{2010SoPh..266..247S}
\begin{barticle}
\bauthor{\bsnm{{Sivaraman}}, \binits{K.R.}},
\bauthor{\bsnm{{Sivaraman}}, \binits{H.}},
\bauthor{\bsnm{{Gupta}}, \binits{S.S.}},
\bauthor{\bsnm{{Howard}}, \binits{R.F.}}:
\byear{2010},
\batitle{{Return Meridional Flow in the Convection Zone from Latitudinal
  Motions of Umbrae of Sunspot Groups}}.
\bjtitle{\solphys}
\bvolume{266},
\bfpage{247}.
\doiurl{https://doi.org/10.1007/s11207-010-9620-6}.
\adsurl{2010SoPh..266..247S}.
\end{barticle}
\endbibitem

\bibitem[\protect\citeauthoryear{{Strugarek}
  et~al.}{2018}]{2018ApJ...863...35S}
\begin{barticle}
\bauthor{\bsnm{{Strugarek}}, \binits{A.}},
\bauthor{\bsnm{{Beaudoin}}, \binits{P.}},
\bauthor{\bsnm{{Charbonneau}}, \binits{P.}},
\bauthor{\bsnm{{Brun}}, \binits{A.S.}}:
\byear{2018},
\batitle{{On the Sensitivity of Magnetic Cycles in Global Simulations of
  Solar-like Stars}}.
\bjtitle{\apj}
\bvolume{863},
\bfpage{35}.
\doiurl{https://doi.org/10.3847/1538-4357/aacf9e}.
\adsurl{2018ApJ...863...35S}.
\end{barticle}
\endbibitem

\bibitem[\protect\citeauthoryear{{Svalgaard} and
  {Kamide}}{2013}]{2013ApJ...763...23S}
\begin{barticle}
\bauthor{\bsnm{{Svalgaard}}, \binits{L.}},
\bauthor{\bsnm{{Kamide}}, \binits{Y.}}:
\byear{2013},
\batitle{{Asymmetric Solar Polar Field Reversals}}.
\bjtitle{\apj}
\bvolume{763},
\bfpage{23}.
\doiurl{https://doi.org/10.1088/0004-637X/763/1/23}.
\adsurl{2013ApJ...763...23S}.
\end{barticle}
\endbibitem

\bibitem[\protect\citeauthoryear{{S{\'y}kora} and
  {Ryb{\'a}k}}{2010}]{2010SoPh..261..321S}
\begin{barticle}
\bauthor{\bsnm{{S{\'y}kora}}, \binits{J.}},
\bauthor{\bsnm{{Ryb{\'a}k}}, \binits{J.}}:
\byear{2010},
\batitle{{Manifestations of the North - South Asymmetry in the Photosphere and
  in the Green Line Corona}}.
\bjtitle{\solphys}
\bvolume{261},
\bfpage{321}.
\doiurl{https://doi.org/10.1007/s11207-009-9483-x}.
\adsurl{2010SoPh..261..321S}.
\end{barticle}
\endbibitem

\bibitem[\protect\citeauthoryear{{Temmer} et~al.}{2006}]{2006A&A...447..735T}
\begin{barticle}
\bauthor{\bsnm{{Temmer}}, \binits{M.}},
\bauthor{\bsnm{{Ryb{\'a}k}}, \binits{J.}},
\bauthor{\bsnm{{Bend{\'\i}k}}, \binits{P.}},
\bauthor{\bsnm{{Veronig}}, \binits{A.}},
\bauthor{\bsnm{{Vogler}}, \binits{F.}},
\bauthor{\bsnm{{Otruba}}, \binits{W.}},
\bauthor{\bsnm{{P{\"o}tzi}}, \binits{W.}},
\bauthor{\bsnm{{Hanslmeier}}, \binits{A.}}:
\byear{2006},
\batitle{{Hemispheric sunspot numbers \{R$_{n}$\} and \{R$_{s}$\} from
  1945-2004: catalogue and N-S asymmetry analysis for solar cycles 18-23}}.
\bjtitle{\aap}
\bvolume{447},
\bfpage{735}.
\doiurl{https://doi.org/10.1051/0004-6361:20054060}.
\adsurl{2006A&A...447..735T}.
\end{barticle}
\endbibitem

\bibitem[\protect\citeauthoryear{{Tokumaru}, {Fujiki}, and
  {Iju}}{2015}]{2015JGRA..120.3283T}
\begin{barticle}
\bauthor{\bsnm{{Tokumaru}}, \binits{M.}},
\bauthor{\bsnm{{Fujiki}}, \binits{K.}},
\bauthor{\bsnm{{Iju}}, \binits{T.}}:
\byear{2015},
\batitle{{North-south asymmetry in global distribution of the solar wind speed
  during 1985-2013}}.
\bjtitle{J. Geophys. Res. (Space Phys.)}
\bvolume{120},
\bfpage{3283}.
\doiurl{https://doi.org/10.1002/2014JA020765}.
\adsurl{2015JGRA..120.3283T}.
\end{barticle}
\endbibitem

\bibitem[\protect\citeauthoryear{{Torrence} and
  {Compo}}{1998}]{1998BAMS..79..61T}
\begin{barticle}
\bauthor{\bsnm{{Torrence}}, \binits{C.}},
\bauthor{\bsnm{{Compo}}, \binits{G.P.}}:
\byear{1998},
\batitle{{A Practical Guide to Wavelet Analysis}}.
\bjtitle{\it Bull. Am. Meteo. Soc.}
\bvolume{79},
\bfpage{61}.
\end{barticle}
\endbibitem

\bibitem[\protect\citeauthoryear{{Usoskin} et~al.}{2017}]{2017yCat..36010109U}
\begin{botherref}
\oauthor{\bsnm{{Usoskin}}, \binits{I.G.}},
\oauthor{\bsnm{{Gallet}}, \binits{Y.}},
\oauthor{\bsnm{{Lopes}}, \binits{F.}},
\oauthor{\bsnm{{Kovaltsov}}, \binits{G.A.}},
\oauthor{\bsnm{{Hulot}}, \binits{G.}},
\oauthor{\bsnm{{Willamo}}, \binits{T.}},
\oauthor{\bsnm{{Usoskin}}, \binits{I.G.}},
\oauthor{\bsnm{{Kovaltsov}}, \binits{G.A.}}:
2017,
{VizieR Online Data Catalog: Monthly numbers of sunspot groups 1749-1996
  (Usoskin+, 2017)}.
\textit{VizieR Online Data Catalog},
J/A+A/601/A109.
\adsurl{2017yCat..36010109U}.
\end{botherref}
\endbibitem

\bibitem[\protect\citeauthoryear{{Vaquero}}{2007}]{2007AdSpR..40..929V}
\begin{barticle}
\bauthor{\bsnm{{Vaquero}}, \binits{J.M.}}:
\byear{2007},
\batitle{{Historical sunspot observations: A review}}.
\bjtitle{\adv}
\bvolume{40},
\bfpage{929}.
\doiurl{https://doi.org/10.1016/j.asr.2007.01.087}.
\adsurl{2007AdSpR..40..929V}.
\end{barticle}
\endbibitem

\bibitem[\protect\citeauthoryear{{Vecchio} et~al.}{2012}]{2012ApJ...749...27V}
\begin{barticle}
\bauthor{\bsnm{{Vecchio}}, \binits{A.}},
\bauthor{\bsnm{{Laurenza}}, \binits{M.}},
\bauthor{\bsnm{{Meduri}}, \binits{D.}},
\bauthor{\bsnm{{Carbone}}, \binits{V.}},
\bauthor{\bsnm{{Storini}}, \binits{M.}}:
\byear{2012},
\batitle{{The Dynamics of the Solar Magnetic Field: Polarity Reversals,
  Butterfly Diagram, and Quasi-biennial Oscillations}}.
\bjtitle{\apj}
\bvolume{749},
\bfpage{27}.
\doiurl{https://doi.org/10.1088/0004-637X/749/1/27}.
\adsurl{2012ApJ...749...27V}.
\end{barticle}
\endbibitem

\bibitem[\protect\citeauthoryear{{Verma}}{1987}]{1987SoPh..114..185V}
\begin{barticle}
\bauthor{\bsnm{{Verma}}, \binits{V.K.}}:
\byear{1987},
\batitle{{On the increase of solar activity in the Southern Hemisphere during
  solar cycle 21}}.
\bjtitle{\solphys}
\bvolume{114},
\bfpage{185}.
\adsurl{1987SoPh..114..185V}.
\end{barticle}
\endbibitem

\bibitem[\protect\citeauthoryear{{Vernova} et~al.}{2004}]{2004SoPh..221..151V}
\begin{barticle}
\bauthor{\bsnm{{Vernova}}, \binits{E.S.}},
\bauthor{\bsnm{{Mursula}}, \binits{K.}},
\bauthor{\bsnm{{Tyasto}}, \binits{M.I.}},
\bauthor{\bsnm{{Baranov}}, \binits{D.G.}}:
\byear{2004},
\batitle{{Long-term longitudinal asymmetries in sunspot activity: Difference
  between the ascending and descending phase of the solar cycle}}.
\bjtitle{\solphys}
\bvolume{221},
\bfpage{151}.
\doiurl{https://doi.org/10.1023/B:SOLA.0000033367.32977.71}.
\adsurl{2004SoPh..221..151V}.
\end{barticle}
\endbibitem

\bibitem[\protect\citeauthoryear{{Vernova} et~al.}{2018}]{2018SoPh..293..158V}
\begin{barticle}
\bauthor{\bsnm{{Vernova}}, \binits{E.S.}},
\bauthor{\bsnm{{Tyasto}}, \binits{M.I.}},
\bauthor{\bsnm{{Baranov}}, \binits{D.G.}},
\bauthor{\bsnm{{Danilova}}, \binits{O.A.}}:
\byear{2018},
\batitle{{Polarity Imbalance of the Photospheric Magnetic Field}}.
\bjtitle{\solphys}
\bvolume{293},
\bfpage{158}.
\doiurl{https://doi.org/10.1007/s11207-018-1382-6}.
\adsurl{2018SoPh..293..158V}.
\end{barticle}
\endbibitem

\bibitem[\protect\citeauthoryear{{Vizoso} and
  {Ballester}}{1990}]{1990A&A...229..540V}
\begin{barticle}
\bauthor{\bsnm{{Vizoso}}, \binits{G.}},
\bauthor{\bsnm{{Ballester}}, \binits{J.L.}}:
\byear{1990},
\batitle{{The north-south asymmetry of sunspots}}.
\bjtitle{\aap}
\bvolume{229},
\bfpage{540}.
\adsurl{1990A&A...229..540V}.
\end{barticle}
\endbibitem

\bibitem[\protect\citeauthoryear{{Waldmeier}}{1957}]{1957ZA.....43..149W}
\begin{barticle}
\bauthor{\bsnm{{Waldmeier}}, \binits{M.}}:
\byear{1957},
\batitle{{Der lange Sonnenzyklus. Mit 3 Textabbildungen}}.
\bjtitle{\zap}
\bvolume{43},
\bfpage{149}.
\adsurl{1957ZA.....43..149W}.
\end{barticle}
\endbibitem

\bibitem[\protect\citeauthoryear{{Waldmeier}}{1971}]{1971SoPh...20..332W}
\begin{barticle}
\bauthor{\bsnm{{Waldmeier}}, \binits{M.}}:
\byear{1971},
\batitle{{The asymmetry of solar activity in the years 1959 1969}}.
\bjtitle{\solphys}
\bvolume{20},
\bfpage{332}.
\doiurl{https://doi.org/10.1007/BF00159763}.
\adsurl{1971SoPh...20..332W}.
\end{barticle}
\endbibitem

\bibitem[\protect\citeauthoryear{{Wang} and
  {Sheeley}}{1991}]{1991ApJ...375..761W}
\begin{barticle}
\bauthor{\bsnm{{Wang}}, \binits{Y.-M.}},
\bauthor{\bsnm{{Sheeley}}, \binits{J.} \bsuffix{N.~R.}}:
\byear{1991},
\batitle{{Magnetic Flux Transport and the Sun's Dipole Moment: New Twists to
  the Babcock-Leighton Model}}.
\bjtitle{\apj}
\bvolume{375},
\bfpage{761}.
\doiurl{https://doi.org/10.1086/170240}.
\adsurl{1991ApJ...375..761W}.
\end{barticle}
\endbibitem

\bibitem[\protect\citeauthoryear{{Watson} et~al.}{2009}]{2009SoPh..260....5W}
\begin{barticle}
\bauthor{\bsnm{{Watson}}, \binits{F.}},
\bauthor{\bsnm{{Fletcher}}, \binits{L.}},
\bauthor{\bsnm{{Dalla}}, \binits{S.}},
\bauthor{\bsnm{{Marshall}}, \binits{S.}}:
\byear{2009},
\batitle{{Modelling the Longitudinal Asymmetry in Sunspot Emergence: The Role
  of the Wilson Depression}}.
\bjtitle{\solphys}
\bvolume{260},
\bfpage{5}.
\doiurl{https://doi.org/10.1007/s11207-009-9420-z}.
\adsurl{2009SoPh..260....5W}.
\end{barticle}
\endbibitem

\bibitem[\protect\citeauthoryear{{Xie}, {Shi}, and
  {Qu}}{2018}]{2018ApJ...855...84X}
\begin{barticle}
\bauthor{\bsnm{{Xie}}, \binits{J.}},
\bauthor{\bsnm{{Shi}}, \binits{X.}},
\bauthor{\bsnm{{Qu}}, \binits{Z.}}:
\byear{2018},
\batitle{{North-South Asymmetry of the Rotation of the Solar Magnetic Field}}.
\bjtitle{\apj}
\bvolume{855},
\bfpage{84}.
\doiurl{https://doi.org/10.3847/1538-4357/aaae68}.
\adsurl{2018ApJ...855...84X}.
\end{barticle}
\endbibitem

\bibitem[\protect\citeauthoryear{{Yeo}, {Krivova}, and
  {Solanki}}{2017}]{2017JGRA..122.3888Y}
\begin{barticle}
\bauthor{\bsnm{{Yeo}}, \binits{K.L.}},
\bauthor{\bsnm{{Krivova}}, \binits{N.A.}},
\bauthor{\bsnm{{Solanki}}, \binits{S.K.}}:
\byear{2017},
\batitle{{EMPIRE: A robust empirical reconstruction of solar irradiance
  variability}}.
\bjtitle{J. Geophys. Res. (Space Phys.)}
\bvolume{122},
\bfpage{3888}.
\doiurl{https://doi.org/10.1002/2016JA023733}.
\adsurl{2017JGRA..122.3888Y}.
\end{barticle}
\endbibitem

\bibitem[\protect\citeauthoryear{{Yi}}{1992}]{1992JRASC..86...89Y}
\begin{barticle}
\bauthor{\bsnm{{Yi}}, \binits{W.}}:
\byear{1992},
\batitle{{The North-South Asymmetry of Sunspot Distribution}}.
\bjtitle{\jrasc}
\bvolume{86},
\bfpage{89}.
\adsurl{1992JRASC..86...89Y}.
\end{barticle}
\endbibitem

\bibitem[\protect\citeauthoryear{{Zaqarashvili}}{2018}]{2018ApJ...856...32Z}
\begin{barticle}
\bauthor{\bsnm{{Zaqarashvili}}, \binits{T.}}:
\byear{2018},
\batitle{{Equatorial Magnetohydrodynamic Shallow Water Waves in the Solar
  Tachocline}}.
\bjtitle{\apj}
\bvolume{856},
\bfpage{32}.
\doiurl{https://doi.org/10.3847/1538-4357/aab26f}.
\adsurl{2018ApJ...856...32Z}.
\end{barticle}
\endbibitem

\bibitem[\protect\citeauthoryear{{Zaqarashvili}
  et~al.}{2010}]{2010ApJ...724L..95Z}
\begin{barticle}
\bauthor{\bsnm{{Zaqarashvili}}, \binits{T.V.}},
\bauthor{\bsnm{{Carbonell}}, \binits{M.}},
\bauthor{\bsnm{{Oliver}}, \binits{R.}},
\bauthor{\bsnm{{Ballester}}, \binits{J.L.}}:
\byear{2010},
\batitle{{Quasi-biennial Oscillations in the Solar Tachocline Caused by
  Magnetic Rossby Wave Instabilities}}.
\bjtitle{\apjl}
\bvolume{724},
\bfpage{L95}.
\doiurl{https://doi.org/10.1088/2041-8205/724/1/L95}.
\adsurl{2010ApJ...724L..95Z}.
\end{barticle}
\endbibitem

\bibitem[\protect\citeauthoryear{{Zaqarashvili}
  et~al.}{2015}]{2015ApJ...805L..14Z}
\begin{barticle}
\bauthor{\bsnm{{Zaqarashvili}}, \binits{T.V.}},
\bauthor{\bsnm{{Oliver}}, \binits{R.}},
\bauthor{\bsnm{{Hanslmeier}}, \binits{A.}},
\bauthor{\bsnm{{Carbonell}}, \binits{M.}},
\bauthor{\bsnm{{Ballester}}, \binits{J.L.}},
\bauthor{\bsnm{{Gachechiladze}}, \binits{T.}},
\bauthor{\bsnm{{Usoskin}}, \binits{I.G.}}:
\byear{2015},
\batitle{{Long-term variation in the Sun's activity caused by magnetic Rossby
  waves in the tachocline}}.
\bjtitle{\apjl}
\bvolume{805},
\bfpage{L14}.
\doiurl{https://doi.org/10.1088/2041-8205/805/2/L14}.
\adsurl{2015ApJ...805L..14Z}.
\end{barticle}
\endbibitem

\bibitem[\protect\citeauthoryear{{Zhang} and
  {Feng}}{2015}]{2015AJ....150...74Z}
\begin{barticle}
\bauthor{\bsnm{{Zhang}}, \binits{J.}},
\bauthor{\bsnm{{Feng}}, \binits{W.}}:
\byear{2015},
\batitle{{Regularity of the North-South Asymmetry of Solar Activity:
  Revisited}}.
\bjtitle{\aj}
\bvolume{150},
\bfpage{74}.
\doiurl{https://doi.org/10.1088/0004-6256/150/3/74}.
\adsurl{2015AJ....150...74Z}.
\end{barticle}
\endbibitem

\bibitem[\protect\citeauthoryear{{Zhao}, {Kosovichev}, and
  {Bogart}}{2014}]{2014ApJ...789L...7Z}
\begin{barticle}
\bauthor{\bsnm{{Zhao}}, \binits{J.}},
\bauthor{\bsnm{{Kosovichev}}, \binits{A.G.}},
\bauthor{\bsnm{{Bogart}}, \binits{R.S.}}:
\byear{2014},
\batitle{{Solar Meridional Flow in the Shallow Interior during the Rising Phase
  of Cycle 24}}.
\bjtitle{\apjl}
\bvolume{789},
\bfpage{L7}.
\doiurl{https://doi.org/10.1088/2041-8205/789/1/L7}.
\adsurl{2014ApJ...789L...7Z}.
\end{barticle}
\endbibitem

\bibitem[\protect\citeauthoryear{{Zirnstein}
  et~al.}{2020}]{2020ApJ...894...13Z}
\begin{barticle}
\bauthor{\bsnm{{Zirnstein}}, \binits{E.J.}},
\bauthor{\bsnm{{Dayeh}}, \binits{M.A.}},
\bauthor{\bsnm{{McComas}}, \binits{D.J.}},
\bauthor{\bsnm{{Sok{\'o}{\l}}}, \binits{J.M.}}:
\byear{2020},
\batitle{{Asymmetric Structure of the Solar Wind and Heliosphere from IBEX
  Observations}}.
\bjtitle{\apj}
\bvolume{894},
\bfpage{13}.
\doiurl{https://doi.org/10.3847/1538-4357/ab8470}.
\adsurl{2020ApJ...894...13Z}.
\end{barticle}
\endbibitem

\bibitem[\protect\citeauthoryear{{Zolotova} and
  {Ponyavin}}{2006}]{2006A&A...449L...1Z}
\begin{barticle}
\bauthor{\bsnm{{Zolotova}}, \binits{N.V.}},
\bauthor{\bsnm{{Ponyavin}}, \binits{D.I.}}:
\byear{2006},
\batitle{{Phase asynchrony of the north-south sunspot activity}}.
\bjtitle{\aap}
\bvolume{449},
\bfpage{L1}.
\doiurl{https://doi.org/10.1051/0004-6361:200600013}.
\adsurl{2006A&A...449L...1Z}.
\end{barticle}
\endbibitem
  
\end{thebibliography}
%

\end{article} 
\end{document}